\documentclass[12pt]{article}
\usepackage{amssymb,amsmath}
\makeatletter

\usepackage{arydshln}

\newcount\comment@nesting

\newenvironment{comment}{%
    \begingroup
    \global\comment@nesting=0
    \def\do##1{\catcode`##1=12\relax}%
    \dospecials
    \catcode`\^^M=\active
    \comment@processline
}{%
    \endgroup
}

\begingroup
\xdef\comment@begincomment{\string\\begin\string\{comment\string\}}
\xdef\comment@endcomment{\string\\end\string\{comment\string\}}
\endgroup

\begingroup
\catcode`\^^M=\active
\def\@temp{\endgroup\def\comment@processline##1^^M}%
\@temp{%
    \def\comment@curline{#1}%
    \let\@next=\comment@processline
    \ifx\comment@curline\comment@endcomment
        \ifnum\comment@nesting=0
            \def\@next{\end{comment}}%
        \else
            \global\advance\comment@nesting by -1
        \fi
    \else
        \ifx\comment@curline\comment@begincomment
            \global\advance\comment@nesting by 1
        \fi
    \fi
    \@next
}

\makeatother

\usepackage[dvipdfmx]{graphicx}
\usepackage{bm}
\usepackage{color}
\usepackage{here}
\usepackage{enumerate}
\usepackage{here}
\usepackage[vcentermath]{youngtab}
\usepackage{subfigure}
\usepackage{tikz}
\usepackage{hyperref}

\newcommand{\Zb}{\mathbb{Z}}

\newcommand{\Cb}{\mathbb{C}}

\newcommand{\Acal}{\mathcal{A}}

\newcommand{\Dcal}{\mathcal{D}}
\newcommand{\Lcal}{\mathcal{L}}

\newcommand{\Ncal}{\mathcal{N}}
\newcommand{\Pcal}{\mathcal{P}}

\DeclareMathOperator*{\Tr}{{\rm Tr}}

\newcommand{\II}{\mathbb{II}}

\numberwithin{equation}{section}

\allowdisplaybreaks[1]

\definecolor{mygreen}{rgb}{0,0.714,0.286}

\begin{document}

\thispagestyle{empty}
\begin{flushright}

\end{flushright}
\vskip1.5cm
\begin{center}
{\Large \bf 
%
%
$SL(2, \Zb)$ dualities of boundary conditions\\
\bigskip
in Abelian
M2-brane SCFTs
}

\vskip1.5cm
Tadashi Okazaki\footnote{tokazaki@seu.edu.cn}

\bigskip
{\it Shing-Tung Yau Center and School of Physics of Southeast University,\\
Yifu Architecture Building, No.2 Sipailou, Xuanwu district, \\
Nanjing, Jiangsu, 210096, China
}

\bigskip
and
\\
\bigskip
Douglas J. Smith\footnote{douglas.smith@durham.ac.uk}

\bigskip
{\it Department of Mathematical Sciences, Durham University,\\
Upper Mountjoy, Stockton Road, Durham DH1 3LE, UK}

\end{center}

\vskip1cm
\begin{abstract}
We propose $SL(2,\mathbb{Z})$ dualities of supersymmetric boundary conditions 
in the three-dimensional supersymmetric field theories describing a semi-infinite M2-brane terminating on M5-branes. 
Specifically, we present dualities of boundary conditions for Abelian (quiver) ADHM theories and circular quiver Chern-Simons matter theories including the ABJM model. For the circular quiver Chern-Simons theories we take boundary conditions breaking a $U(1)_1 \times U(1)_{-1}$ gauge group to its diagonal subgroup which is decoupled. 
This can be generalized to break $U(1)_k \times U(1)_{-k}$, leaving a $\Zb_k$ gauge theory.
We find matching of the 't Hooft anomalies and supersymmetric half-indices for all the proposed dual boundary conditions.
\end{abstract}
\newpage
\setcounter{tocdepth}{3}
\tableofcontents

\section{Introduction}
\label{sec_intro}

In this paper we study BPS boundary conditions in M2-brane SCFTs, 
that is the 3d $\mathcal{N}\ge 4$ supersymmetric field theories which appear at low energy on the world-volume of a stack of M2-branes. 
There are various conjectural dual descriptions of a stack of M2-branes, 
including $U(N)$ ADHM theories \cite{deBoer:1996mp,deBoer:1996ck}, $U(N)_k\times U(N)_{-k}$ ABJM theories \cite{Aharony:2008ug} 
and circular quiver Chern-Simons theories \cite{Gaiotto:2008sd,Imamura:2008nn,Imamura:2008dt}. 
Brane configurations in Type IIB string theory \cite{Hanany:1996ie,Gaiotto:2008ak} are useful 
to study the action of dualities of the 3d $\mathcal{N}\ge 4$ supersymmetric field theories. 
Such dualities can be generalized by introducing various defect operators residing in the theories.

The case of boundaries for M2-brane theories is a much explored topic although not fully understood. A natural interpretation is in terms of configurations of M2-branes ending on M5-branes. For multiple M2-branes the Basu-Harvey equation \cite{Basu:2004ed} describes a generalization of the Nahm equation with the M2-branes creating an M5-brane via a fuzzy funnel, and also admits solutions corresponding to M2-branes between two M5-branes \cite{Nogradi:2005yk}. Properties of these solutions have been explored further in terms of the Basu-Harvey equation and in both BLG and ABJM theories \cite{Berman:2005re, Berman:2006eu, Hanaki:2008cu, Nastase:2009ny, Nosaka:2012tq}. In particular, the Basu-Harvey equation was a key ingredient in the derivation of the BLG model of M2-branes \cite{Bagger:2006sk, Bagger:2008se, Gustavsson:2007vu} and is also present in the ABJM model \cite{Aharony:2008ug}. Other work has focussed on the allowed boundary conditions and the derivation of a boundary action to describe the self-dual strings arising as the boundary of M2-branes within M5-branes \cite{Berman:2009kj, Chu:2009ms, Berman:2009xd, Faizal:2011cd, Hori:2013ewa, Hosomichi:2014rqa, Armoni:2015jsa, Niarchos:2015lla, Faizal:2016skd}. It is also interesting to study topological twists of the systems of M2-branes and M5-branes labelled M-strings \cite{Haghighat:2013gba} which can give rise to supergroup WZW theories \cite{Okazaki:2015fiq, Okazaki:2016pne}.

In this paper we propose $SL(2,\mathbb{Z})$ dualities of the chiral supersymmetric boundary conditions in the Abelian M2-brane SCFTs. 
The basic boundary conditions in the ADHM (quiver) theories, 
where the vector multiplet and hypermultiplets obey the same types of $\mathcal{N}=(0,4)$ Neumann or Dirichlet boundary conditions, 
can be realized in Type IIB string theory by taking periodic D3-branes in the brane setup \cite{Chung:2016pgt,Hanany:2018hlz}. 
The mirror transformation of the boundary conditions in these theories can be realized as the $S$ transformation of the $SL(2,\mathbb{Z})$ duality action in Type IIB string theory. 
The dualities of these boundary conditions in the Abelian gauge theories 
can be viewed as the circular versions of those in the 3d $\mathcal{N}=4$ linear quiver gauge theories \cite{Okazaki:2019bok}. 
When the opposite types of boundary conditions for the vector multiplet and the adjoint hypermultiplet are chosen, 
we find two types of self-mirror boundary conditions. 
They are expected to be realized when one introduces the boundary $(-1,1)$ and $(1,1)$ $5$-branes 
which are invariant under the $S$ transformation. 
The ADHM (quiver) theories are also dual to the circular quiver Chern-Simons theories upon the $STS$ transformation. We claim that it relates to new types of boundary conditions which partially break the gauge group.
The latter theories contain $\mathcal{N}=2$ $U(1)_{k}\times U(1)_{-k}$ vector multiplets with opposite Chern-Simons levels rather than $\mathcal{N}=4$ vector multiplets. 
For $k=1$, we choose the $\Ncal = (0,2)$ boundary condition \cite{Okazaki:2013kaa} that partially breaks the gauge group down to its diagonal subgroup which is decoupled. 
For general $k$ it can be generalized to the boundary condition which breaks $U(1)_k \times U(1)_{-k}$ gauge group, leaving a $\Zb_k$ gauge theory together with the decoupled diagonal subgroup. 
We argue that these boundary conditions in the Abelian circular quiver Chern-Simons theories are dual to the basic boundary conditions in the ADHM (quiver) theories 
by checking precise matching of the boundary 't Hooft anomalies and the half-indices.

\subsection{Structure}
The organization of the paper is as follows. 
In section \ref{sec_bcM2SCFT} we summarize the brane setup in M-theory, Type IIA string theory and Type IIB string theory. 
We then study the BPS boundary conditions in the M2-brane SCFTs which can preserve $\mathcal{N}=(0,4)$ supersymmetry. 
In section \ref{sec_hindex} we examine the half-indices for the BPS boundary conditions. 
In section \ref{sec_Dualities} we propose the $SL(2,\mathbb{Z})$ dualities of the BPS boundary conditions.
In appendix~\ref{ABJM_U1k_VMbc} we comment on some potential alternative boundary conditions for $U(1)_k \times U(1)_{-k}$ ABJM theory but these do not fit with the dualities discussed in the main body of the paper.

\subsection{Future works}

\begin{itemize}

\item Our results indicate that the brane construction \cite{Chung:2016pgt,Hanany:2018hlz} 
of the half-BPS boundary conditions in 3d $\mathcal{N}\ge 4$ supersymmetric gauge theories 
in terms of the NS5- and D5-branes can be generalized by introducing more general $(p,q)$ 5-branes. 
It would be nice to further understand the resulting boundary conditions in $\mathcal{N}=2$ or $3$ supersymmetric gauge theories 
and the dualities using the brane construction \cite{deBoer:1997ka,Aharony:1997ju,Kitao:1998mf,Giveon:2008zn}. 

\item The dualities of the boundary conditions can be further generalized by introducing line defect operators. 
In the absence of the boundary conditions the line defect half-BPS indices in the ADHM theories were studied in \cite{Hayashi:2024jof}. 
Also the dualities of the line defects were examined for 3d $\mathcal{N}\ge 4$ SCFTs in 
\cite{Assel:2015oxa,Dimofte:2019zzj,Dey:2021jbf,Nawata:2021nse,Hayashi:2025guk}. 
It would be interesting to figure out the dualities of the configurations involving both boundaries and lines by studying the line defect half-indices for the M2-brane SCFTs. 

\item While we have mainly focused on the chiral supersymmetric boundary conditions in this work, 
the 3d $\mathcal{N}\ge 4$ supersymmetric field theories also have non-chiral $\mathcal{N}=(2,2)$ half-BPS boundary conditions \cite{Bullimore:2016nji,Chung:2016pgt,Okazaki:2020lfy,Okazaki:2023kpq}. 
In particular, the duality transformation under mirror symmetry of a collection of half-indices for the exceptional Dirichlet boundary conditions \cite{Bullimore:2016nji}
is described by the triangular matrix obtained from the elliptic stable envelope \cite{Aganagic:2016jmx}, 
as demonstrated for the Abelian linear quiver gauge theories \cite{Okazaki:2020lfy}. 
It would be intriguing to examine the mirror transformation of the $\mathcal{N}=(2,2)$ half-BPS boundary conditions in the circular quiver setup. 

\item The $\mathcal{N}=(0,4)$ half-BPS boundary conditions in 3d $\mathcal{N}=4$ supersymmetric non-Abelian gauge theories 
admit the Nahm pole boundary conditions \cite{Chung:2016pgt,Gaiotto:2019jvo}. 
For such singular boundary conditions, the Higgsing procedure \cite{Gaiotto:2012xa} would be useful to compute the half-indices, 
as demonstrated for the 4d cases \cite{Gaiotto:2019jvo,Hatsuda:2024lcc,Hatsuda:2025yzp}. 

\item Deriving the exact closed-form formulae of the half-indices would be highly desirable in further study. 
In particular, the Neumann half-indices for the ADHM (quiver) theories are given by intriguing matrix integrals. 
They reduce to the refined Hilbert series \cite{Benvenuti:2010pq,Hanany:2012dm,Cremonesi:2013lqa} for the Higgs/Coulomb branches for the ADHM theories 
in the Higgs/Coulomb limits. 
On the other hand, in the topologically twisted limits, they can be viewed as the vacuum characters of the associated boundary VOAs \cite{Costello:2018fnz}, 
which generalize the unflavored Schur indices for $\mathcal{N}=4$ SYM and $\mathcal{N}=2$ circular quiver gauge theories 
\cite{Bourdier:2015wda,Bourdier:2015sga}.

\item The holographic dual geometries for M2-branes intersecting with M5-branes in the near-horizon limit of the M2-branes were investigated 
in \cite{DHoker:2008lup,DHoker:2008rje,DHoker:2008wvd,DHoker:2009lky,DHoker:2009wlx,Bachas:2013vza,Bena:2023rzm,Bena:2024dre}. 
They give rise to intriguing examples for studying various phenomena of ETW branes as a part of the $AdS_4$ is now cut off by an ETW brane with $AdS_3$ factor. 
It would be an interesting problem to examine the spectrum in the dual gravity side. 
The KK modes on the dual geometries should account for our result of the large $N$ limit of the half-index. 
Furthermore, the finite $N$ correction that would encode the spectrum of the giant gravitons \cite{McGreevy:2000cw} in the dual geometries 
will be obtained by performing the giant graviton expansions \cite{Arai:2019xmp,Arai:2020qaj,Gaiotto:2021xce} of the half-indices. 
It should generalize the giant graviton expansions of the Coulomb/Higgs indices in \cite{Gaiotto:2021xce,Beccaria:2023sph,Hayashi:2024aaf}. 

\end{itemize}

\section{Boundary conditions for M2-brane SCFTs}
\label{sec_bcM2SCFT}

\subsection{M-theory brane setup}
We consider the following brane configuration in M-theory
\begin{align}
\label{M_branesetup}
\begin{array}{c|ccccccccccc}
     & 0 & 1 & 2 & 3 & 4 & 5 & 6 & 7 & 8 & 9 & 10 \\ \hline
N \; \textrm{M2} & \circ & \circ & \circ & & & & & & & & \\
\textrm{TN}_{k = 1} & & & & & & & \times & \times & \times & \times & \\
\textrm{TN}_l & & & & \times & \times & \times & & & & & \times \\ \hline
\textrm{M5$'$} & \circ & \circ & & & & & \circ &\circ & \circ & \circ & \\
\widetilde{\textrm{M5$'$}} & \circ & \circ & & \circ & \circ & \circ & & & & &\circ
\end{array}
\end{align}
Here $\circ$ stands for the directions in which branes are extended. 
$\textrm{TN}_l$ refers to an $l$-centered Taub-NUT space which is asymptotically $\Cb^2/\Zb_l$ in the directions marked with $\times$.
In general, we would have an asymptotic $\Cb^2/\Zb_k \times \Cb^2/\Zb_l$ orbifold 
but in this paper we focus only on the case of $k = 1$ giving a $\Cb^2/\Zb_l$ orbifold.
We consider the configurations where the $N$ semi-infinite M2-branes terminate on the M5-branes.
In the IR limit we can simply consider $\textrm{TN}_l$ to be $\Cb^2/\Zb_l$. 
So, we could choose to ignore the $\textrm{TN}_1$. 
Also, for $l=1$ we could choose to ignore the $\textrm{TN}_l$. In the case of $l=1$, by either keeping or ignoring both $\textrm{TN}_1$s, 
we see that the M5$'$ and $\widetilde{\textrm{M5}'}$ are equivalent up to relabelling of coordinates $345(10) \leftrightarrow 6789$ 
but there is no such symmetry for $l > 1$. 
The M5$'$- and $\widetilde{\textrm{M5}'}$-branes provide us with boundary conditions in the effective theories of a stack of M2-branes. 
As we will see below, they can preserve chiral $\mathcal{N}=(0,4)$ supersymmetry. 

In the case of an M5$'$-brane we have a configuration of $N$ M2-branes 
ending on the M5$'$-brane with a transverse $l$-centered Taub-NUT background which can be understood as arising from $l$ KK monopoles. 
This then gives rise to the $N$ M2-brane boundary theory on a $\Cb^2/\Zb_l$ background.
Note that when the M2-branes are stretched between a pair of M5$'$-branes, 
one finds the M-strings \cite{Haghighat:2013gba,Haghighat:2013tka}. 
We expect that they can be addressed by generalizing the above setup 
in such a way that the dual brane configuration in Type IIB string theory contains the brane box model \cite{Hanany:2018hlz}. 
But we defer discussions with this regard to future work. 

If we have an $\widetilde{\textrm{M5}'}$-brane instead, it is wrapped on the $l$-centered Taub-NUT space. 
The configuration is T-dual to the system of an $\widetilde{\textrm{D4}'}$-brane intersecting with $l$ D6-branes (see (\ref{IIA_branesetup_ADHM})). 
It contains the chiral fermions arising from the open fundamental strings that stretch between the $\widetilde{\textrm{D4}'}$ and D6-branes. 
They give a realization of the affine Kac-Moody algebra on the two-dimensional space (see e.g. \cite{Itzhaki:2005tu,Dijkgraaf:2007sw,Dijkgraaf:2007fe,Tan:2008wp,Witten:2009at,Ohlsson:2012yn,Lambert:2018mfb,Gustavsson:2022jpo}). 
We further consider a stack of $N$ M2-branes intersecting with the $\widetilde{\textrm{M5}'}$-brane on the Taub-NUT space. 

While directly studying M2-branes and their boundary conditions is not straightforward, we will instead focus mostly on the worldvolume field theories arising on D2-branes when these M-theory brane configurations are reduced to Type IIA, or on D3-branes in Type IIB after a further T-duality. This corresponds to studying boundary conditions in $U(N)$ ADHM theories.
Further dualities lead to ADHM quiver theories and we also explore the relation to ABJM theories.
We also note that there is another interesting family of boundary conditions, preserving $\mathcal{N}=(2,2)$ supersymmetry, 
which is distinguished from the setup (\ref{M_branesetup}). 
After a topological twist we expect M2-branes with boundary to be described by theories with supergroup symmetry, 
such as we previously explored in \cite{Okazaki:2015fiq,Okazaki:2016pne}. 
On the other hand, a topological twist should correspond to taking a special limit of fugacities in the half-indices. 
It would be interesting to understand better how supergroups are manifest in this way from studying the half-indices.

\subsection{$U(N)$ ADHM boundary conditions}
In order to see how the brane setup (\ref{M_branesetup}) can be interpreted in gauge theory, 
we first compactify M-theory along $x^{10}$. Compactification transverse to the $\textrm{TN}_1$ background in the $x^6$, $x^7$, $x^8$, $x^9$ directions produces a $\textrm{TN}_1$ background again but here we choose to label this as a unit charge KK-monopole, whereas $l$ D6-branes arise from compactifying the $x^{10}$ isometry direction of the $\textrm{TN}_l$ background.
One finds
\begin{equation}
\label{IIA_branesetup_ADHM}
\begin{array}{c|cccccccccc}
     & 0 & 1 & 2 & 3 & 4 & 5 & 6 & 7 & 8 & 9 \\ \hline
N \; \textrm{D2} & \circ & \circ & \circ & & & & & & & \\
\textrm{KK} & \circ & \circ & \circ & \circ & \circ & \circ &  &  &  &  \\
l \; \textrm{D6} & \circ & \circ & \circ & & & & \circ & \circ & \circ & \circ \\ \hline
\textrm{NS5$'$} & \circ & \circ & & & & & \circ &\circ & \circ & \circ \\
\widetilde{\textrm{D4$'$}} & \circ & \circ & & \circ & \circ & \circ & & & & 
\end{array}
\end{equation}
Furthermore, upon T-duality along $x^6$, the configuration maps to the following brane configuration in Type IIB string theory
\begin{align}
\label{IIB_branesetup_ADHM}
\begin{array}{c|cccccccccc}
     & 0 & 1 & 2 & 3 & 4 & 5 & 6 & 7 & 8 & 9 \\ \hline
N \; \textrm{D3} & \circ & \circ & \circ & & & & \circ & & & \\
\textrm{NS5} & \circ & \circ & \circ & \circ & \circ & \circ & & & & \\
l \; \textrm{D5} & \circ & \circ & \circ & & & & & \circ & \circ & \circ \\ \hline
\textrm{NS5$'$} & \circ & \circ & & & & & \circ &\circ & \circ & \circ \\
\widetilde{\textrm{D5$'$}} & \circ & \circ & & \circ & \circ & \circ & \circ & & & 
\end{array}
\end{align}
where the $x^6$ direction is taken to be compactified.

Here we introduce either the NS5$'$ or the $\widetilde{\textrm{D5$'$}}$ ($\widetilde{\textrm{D4$'$}}$) brane to impose different boundary conditions on the 3d theory. 
In terms of the two 16-component spinors $\epsilon_L$ and $\epsilon_R$ 
which have the same spacetime chirality $\Gamma_{01234566789}\epsilon_{L/R} = \epsilon_{L/R}$ in Type IIB, the bulk D3, NS5 and D5 branes impose the projection conditions
\begin{align}
    \epsilon_R & = -\Gamma_{0126}\epsilon_L \\
    \epsilon_L & = \Gamma_{6789} \epsilon_L
\end{align}
preserving 8 supercharges, i.e.\ $\mathcal{N} = 4$ supersymmetry in the 3d theory.
In the case of either the NS5$'$-brane or the $\widetilde{\textrm{D5$'$}}$-brane imposing boundary conditions on the 3d theory 
we have the additional constraint $\Gamma_{01}\epsilon_L = -\epsilon_L$ so we have 4 supercharges preserved in both cases, 
and on the 2d boundary we have chiral $\mathcal{N} = (0,4)$ supersymmetry.

Since for $l = 1$ the $U(N)-[1]$ ADHM theory is dual to the $U(N)_1 \times U(N)_{-1}$ ABJM theory, we know that there is supersymmetry enhancement to $\Ncal =8$ in the bulk. Potentially this could result in an enhancement of the supersymmetry in the presence of a boundary up to $\Ncal = (4, 4)$. However, we do not analyze this potential enhancement in this paper.

Note that more generally we could introduce a boundary $(p, q)'$ 5-brane spanning the $01789$ directions and at an angle (dependent on $p$, $q$ and $\tau$) in the $26$-plane or a boundary $\widetilde{(p, q)}'$ 5-brane spanning the $01345$ directions and again at an angle in the $26$-plane. 
In general this allows two different orientations of boundary $(p, q)$ 5-branes with the exception of the NS5-brane or D5-brane 
since the D5$'$-brane would just be the D5-brane spanning the $012789$ directions 
and similarly the $\widetilde{\textrm{NS5$'$}}$-brane would just be the NS5-brane spanning the $012345$ directions.

The $N$ M2-branes with the transverse Taub-NUT space map to $N$ D3-branes together with a single NS5-brane and $l$ D5-branes. 
This gives rise to the 3d $\Ncal = 4$ $U(N)$ ADHM theory, 
a gauge theory with a vector multiplet (VM, $\Phi$) 
an adjoint hypermultiplet $(X, Y)$ and $l$ fundamental hypermultiplets $(I, J)$ \cite{deBoer:1996ck}. 
In $\mathcal{N} = 2$ notation the field content of the $U(N)$ ADHM theory with $l$ flavors is
\begin{align}
\label{UN_l_ADHM_charges}
\begin{array}{c|c|c|c|c|c|c|c}
& U(N) & U(1)_{u_I}^l & SU(l) & U(1)_x & U(1)_t & U(1)_z & U(1)_R \\ \hline
\textrm{VM} & {\bf Adj} & {\bf 0} & {\bf 1} & 0 & 0 & 0 & 0 \\
\Phi & {\bf Adj} & {\bf 0} & {\bf 1} & 0  & -2 & 0 & 1 \\
X & {\bf Adj} & {\bf 0} & {\bf 1} & 1 & 1 & 0 & \frac{1}{2} \\
Y & {\bf Adj} & {\bf 0} & {\bf 1} & -1 & 1 & 0 & \frac{1}{2} \\
I_{\alpha} & {\bf N} & {\bf 0} & {\bf l} & 0 & 1 & 0 & \frac{1}{2} \\
J_{\alpha} & {\bf \overline{N}} & {\bf 0} & {\bf \overline{l}} & 0 & 1 & 0 & \frac{1}{2} \\
\hline
\widetilde{\Gamma}_{I, I+1} & {\bf N} & (\underbrace{0,\cdots,0}_{I-1}, 1, -1, 0,\cdots, 0) & (+1)_I & 0 & 0 & 1 & 0 \\
\widetilde{\eta} & {\bf 1} & (1, 0, \ldots, 0) & {\bf 1} & -1 & 0 & 0 & 0
\end{array}
\end{align}
where for later convenience we have listed the charges of 2d Fermi multiplets $\widetilde{\Gamma}_{I, I+1}$ and $\widetilde{\eta}$, and since we have a cyclic property we identify $u_{l+1}$ with $u_1$. We explain the labeling of these global symmetries and charges in Section~\ref{Sec_ADHM_NNN_Anom}.
While the bulk theory has an $SU(l)$ flavor symmetry, classically the dual theories we discuss later only exhibit a $U(1)^{l-1}$ symmetry which we label with factors $U(1)_{y_{\alpha}}$ with a constraint on the field strengths $\sum_{\alpha = 1}^l y_{\alpha} = 0$ and on the fugacities used in the half-indices this becomes $\prod_{\alpha = 1}^l y_{\alpha} = 1$. The Fermi $\widetilde{\Gamma}_{I, I+1}$ has charge $+1$ under $U(1)_{y_I}$ so with boundary conditions which introduce the Fermis $\widetilde{\Gamma}_{I, I+1}$ we also see only $U(1)^{l-1}$ rather than $SU(l)$. When checking matching of half-indices later we mostly ignore this flavor symmetry for simplicity and also much of the focus is on the case $l = 1$ where there is anyway no symmetry, so we will largely ignore it with some comments in the text about matching of anomalies.

The additional $5$-branes, i.e. NS5$'$-branes and $\widetilde{\textrm{D5$'$}}$-branes in the brane configuration (\ref{IIB_branesetup_ADHM}) 
can be understood as $\mathcal{N}=(0,4)$ half-BPS boundary conditions in the 3d $\mathcal{N}=4$ gauge theory \cite{Chung:2016pgt}. 
Introducing the NS5$'$-brane provides a boundary for the 3d $U(N)$ ADHM theory 
preserving $\Ncal = (0,4)$ supersymmetry with Neumann boundary condition $\mathcal{N}$ for the vector multiplet and 
with Neumann boundary conditions $N$ for the hypermultiplets \cite{Chung:2016pgt}. 
In $\Ncal = 2$ language this is realized as Neumann boundary conditions for VM, $X, Y, I, J$ and Dirichlet boundary condition for $\Phi$.

Instead, for $N = 1$, introducing the $\widetilde{\textrm{D5$'$}}$-brane would give Dirichlet boundary condition $\mathcal{D}$ for the vector multiplet 
and Dirichlet boundary condition $D$ for the hypermultiplets \cite{Chung:2016pgt}. 
In $\Ncal = 2$ language this is Dirichlet boundary conditions for VM, $X, Y, I, J$ and Neumann boundary condition for $\Phi$. 
However, for $N > 1$ the $\widetilde{\textrm{D5$'$}}$-brane would give a singular boundary condition involving the Nahm pole. 
We do not study the case of such Nahm pole boundary conditions here, leaving this interesting topic for future work. We now look at the gauge and 't Hooft anomalies for these ADHM theories with Neumann or Dirichlet boundary conditions.

\subsubsection{ADHM theory with $(\mathcal{N},N,N)+\widetilde{\Gamma} + \widetilde{\eta}$}
\label{Sec_ADHM_NNN_Anom}
Consider the $U(N)$ ADHM theory with $l$ fundamental flavors in terms of $\Ncal = 2$ superfields, 
where the $U(N)$ vector multiplet obeys $\mathcal{N}=(0,2)$ Neumann boundary condition $\mathcal{N}$ with the adjoint chiral 
$\Phi$ subject to $\mathcal{N}=(0,2)$ Dirichlet boundary condition $D$, 
the adjoint hypers $(X, Y)$ subject to $\mathcal{N}=(0,2)$ Neumann boundary conditions $N$ 
and the fundamental hypers $(I, J)$ satisfy $\mathcal{N}=(0,2)$ Neumann boundary conditions $N$ \cite{Okazaki:2013kaa}.
We expect such boundary conditions to arise in the brane configuration where we introduce the NS5$'$-brane \cite{Chung:2016pgt}.
We can easily calculate the gauge and 't Hooft anomalies as follows \cite{Dimofte:2017tpi},
\begin{align}
\label{bdy_UN_1_ADHM_anom_NDNN}
\Acal_{\mathcal{N},N,N} = & \underbrace{N \Tr(s^2) - \Tr(s)^2 + \frac{N^2}{2}r^2}_{\textrm{VM}, \; \Ncal} + \underbrace{\left( N \Tr(s^2) - \Tr(s)^2 + 2N^2t^2 \right)}_{\Phi, \; D}
\nonumber \\
 & - \underbrace{\left( 2N \Tr(s^2) - 2\Tr(s)^2 + N^2 x^2 + N^2\left( t - \frac{1}{2}r \right)^2 \right)}_{(X, Y), \; N}
\nonumber \\
 & - \underbrace{\left( l\Tr(s^2) + N\Tr(y^2) + Nl\left( t - \frac{1}{2}r \right)^2 \right) }_{(I, J), \; N}
  \nonumber \\
  = & -l\Tr(s^2) - N\Tr(y^2) - N^2 x^2
  \nonumber \\
  &  + N(N-l)t^2 + N(N+l)tr + \frac{N(N - l)}{4}r^2 \; .
\end{align}
Here and below we use notation for field strengths $x, t, z$ corresponding to $U(1)_x$, $U(1)_t$, $U(1)_z$, $r$ for $U(1)_R$, $s$ for $U(N)$, $u_I$ for the $U(1)^l_{u_I}$ and $y$ for the $SU(l)$ flavor symmetry. When we later introduce the half-indices we use similar notation for the fugacities.

The gauge anomaly can be cancelled if the $l$ 2d Fermi multiplets $\widetilde{\Gamma}_{I, I+1}$ are introduced. 
In fact, such 2d Fermi multiplets appear from the fluctuations of open strings between $N$ D3-branes and $l$ D5-branes at the intersection 
with the NS5$'$-brane in the Type IIB brane configuration \cite{Hanany:2018hlz}. These Fermis have charges $(+1, -1)$ under $U(1)_{u_I} \times U(1)_{u_{I+1}}$ which can be understood as being related to the left and right of the $I$th D5-brane. As we have a cyclic construction the leftmost $U(1)_{u_1}$ and rightmost $U(1)_{u_{l+1}}$ are identified.
These Fermis also have charge $+1$ under $U(1)_{y_I}$.
Introducing field strengths $u_I$ for the $U(1)_{u_I}^l$ group, the Fermis $\widetilde{\Gamma}_{I, I+1}$ give a contribution
\begin{align}
    \Acal_{\widetilde{\Gamma}} & = l \Tr(s^2) + 2l\Tr(s) z + lNz^2 + N\sum_{I=1}^l \left( u_I - u_{I+1} + y_I \right)^2
\end{align}
to the anomaly. Note that none of these Fermis are charged under the diagonal subgroup of $U(1)_{u_I}^l$ so the global symmetry is really $U(1)^{l-1}$.

In addition, from the intersection of the NS5-brane and NS5$'$-brane 
we would get a bideterminant Fermi $\widetilde{\eta}$ also charged under a $U(1)_{\tilde{u}}$ and $U(1)_x$ \cite{Hanany:2018hlz}.
Here bideterminant means in the determinant representation of the gauge group arising from D3-branes 
on one side of the NS5-brane and inverse determinant of the gauge group arising from D3-branes on the other side of the NS5-brane. 
However, we have a circular configuration so that these two gauge groups are in fact the same gauge group, 
hence the Fermi is in the trivial representation giving anomaly contribution
\begin{align}
    \Acal_{\widetilde{\eta}} & = (\tilde{u} - x)^2 \; .
\end{align}
Together with the $U(1)^{l-1}$ global symmetry this results in a global $U(1)^l$ symmetry and one way to label this is as $U(1)_{u_I}^l$ where we identify $U(1)_{\tilde{u}}$ and $U(1)_{u_1}$, i.e.\ $\tilde{u} \equiv u_1$,
resulting in a total anomaly
\begin{align}
\label{bdy_UN_l_ADHM_anom_NDNN_Gamma}
\Acal_{\mathcal{N},N,N + \widetilde{\Gamma} + \widetilde{\eta}} = & N\sum_{I=1}^l \left( u_I - u_{I+1} \right)^2 + lNz^2 + 2N\sum_{I=1}^l \left( u_I - u_{I+1} \right)y_I - (N^2 - 1) x^2
  \nonumber \\
  &  + u_1^2 - 2 u_1 x + N(N-l)t^2 + N(N+l)tr + \frac{N(N - l)}{4}r^2 \; ,
\end{align}
where we have also included a background FI term $-2l\Tr(s) z$ and noting that when focussing on $U(1)^{l-1}$ rather than $SU(l)$, $\Tr(y^2) = \sum_{I = 1}^l y_I^2$.
We refer to the above quantum mechanically consistent $\mathcal{N}=(0,4)$ boundary condition as $(\mathcal{N},N,N)+\widetilde{\Gamma} + \widetilde{\eta}$.

In the Abelian case, $N=1$, we have
\begin{align}
\label{bdy_U1_l_ADHM_anom_NDNN_Gamma}
\Acal = & \sum_{I=1}^l \left( u_I - u_{I+1} \right)^2 + lz^2 + 2\sum_{I=1}^l \left( u_I - u_{I+1} \right)y_I + u_1^2 - 2 u_1 x
  \nonumber \\
  &  + (1-l)t^2 + (1+l)tr + \frac{1 - l}{4}r^2. 
\end{align}

Alternatively, in the case of $l = 1$ we have anomaly
\begin{align}
\label{bdy_UN_1_ADHM_anom_NDNN_Gamma}
\Acal = & Nz^2 - (N^2 - 1)x^2 + u_1^2 - 2 u_1 x
  \nonumber \\
  &  + N(N-1)t^2 + N(N+1)tr + \frac{N(N-1)}{4}r^2 \; .
\end{align}

If we consider the special case of $N = l = 1$ we have
\begin{align}
\label{bdy_U1_1_ADHM_anom_NDNN_Gamma_eta}
\Acal = & z^2 + u_1^2 - 2u_1x + 2tr \; .
\end{align}

\subsubsection{ADHM theory with $(\mathcal{D},D,D)$}
Consider the $\mathcal{N}=(0,4)$ boundary condition $(\mathcal{D},D,D)$ in the $U(N)$ ADHM theory with $l$ flavors, 
where the $U(N)$ vector multiplet obeys Dirichlet boundary condition $\mathcal{D}$, 
the adjoint hyper is subject to Dirichlet boundary condition $D$ 
and the charged hyper satisfies Dirichlet boundary condition $D$.
We expect such boundary conditions to arise in the brane configuration where we introduce $N$ $\widetilde{\textrm{D5$'$}}$-branes \cite{Chung:2016pgt}.
Note that $\widetilde{\textrm{D5$'$}}$-branes can lead to Nahm pole boundary conditions, and in particular this is the case if there are fewer $\widetilde{\textrm{D5$'$}}$-branes than D3-branes \cite{Gaiotto:2019jvo}. 
We are considering only the special case where each D3-brane ends on a separate $\widetilde{\textrm{D5$'$}}$-brane 
so that the particular Nahm boundary conditions become standard Dirichlet boundary conditions.
The anomaly is
\begin{align}
\label{bdy_UN_l_ADHM_anom_DDD}
\Acal_{\mathcal{D},D,D} = & -\underbrace{\left( N \Tr(u^2) - \Tr(u)^2 + \frac{N^2}{2}r^2 \right)}_{\textrm{VM}, \; \Dcal} - \underbrace{\left( N \Tr(u^2) - \Tr(u)^2 + 2N^2t^2 \right)}_{\Phi, \; N}
\nonumber \\
 & + \underbrace{\left( 2N \Tr(u^2) - 2\Tr(u)^2 + N^2 x^2 + N^2\left( t - \frac{1}{2}r \right)^2 \right)}_{(X, Y), \; D}
\nonumber \\
 & + \underbrace{\left( l\Tr(u^2) + N \Tr(y^2) + lN\left( t - \frac{1}{2}r \right)^2 \right) }_{(I, J), \; D}
 - \underbrace{2lz\Tr(u)}_{FI}
  \nonumber \\
 = & l\Tr(u^2) - 2lz\Tr(u) + N^2 x^2 + N\Tr(y^2) - N(N-l)t^2 
 \nonumber\\
 &- N(N+l)tr - \frac{N(N-l)}{4}r^2 \; .
\end{align}

In the Abelian case we have 
\begin{align}
\label{bdy_U1_l_ADHM_anom_DDD}
\Acal_{\mathcal{D},D,D} = & l u^2 - 2lzu + x^2 + \Tr(y^2) + (l-1)t^2 - (l+1)tr + \frac{l-1}{4}r^2 \; ,
\end{align}
while in the special case of $N = l = 1$ we have anomaly
\begin{align}
\label{bdy_U1_1_ADHM_anom_DNDD}
\Acal_{\mathcal{D},D,D} = & u^2 - 2uz + x^2 - 2tr \; .
\end{align}

\subsection{ADHM quiver boundary conditions}

Note that the $SL(2, \Zb)$ transformations in Type IIB string theory map between different $(p, q)$ 5-branes 
where the NS5 is a $(1, 0)$ 5-brane while the D5-brane is a $(0, 1)$ 5-brane. 
Specifically, under a transformation given by $M = \left( \begin{array}{cc} a & b \\ c & d \end{array} \right) \in SL(2, \Zb)$ where the complexified coupling $\tau$ is mapped to
\begin{align}
    \tau' & = \frac{a \tau + b}{c \tau + d} \; ,
\end{align}
a $(p, q)$ 5-brane is mapped to a $(p', q')$ 5-brane where
\begin{align}
    (p' \; q') & = (p \; q) M^{-1} = (p \; q) \left( \begin{array}{cc} d & -b \\ -c & a \end{array} \right) \; .
\end{align}
In particular, if we define
\begin{align}
    S & = \left( \begin{array}{cc} d & -b \\ -c & a \end{array} \right) = \left( \begin{array}{cc} 0 & -1 \\ 1 & 0 \end{array} \right) \; ,\\
    T & = \left( \begin{array}{cc} d & -b \\ -c & a \end{array} \right) = \left( \begin{array}{cc} 1 & -1 \\ 0 & 1 \end{array} \right), 
\end{align}
then we see that $S$ maps $\tau \to \frac{-1}{\tau}$, NS5 $\to$ $(0, -1)$ and D5 $\to$ NS5 while $T$ maps $\tau \to \tau + 1$, NS5 $\to$ $(1, -1)$ and D5 $\to$ D5.
Note that the $(0, -1)$ 5-brane is an anti-D5-brane but this differs from a D5-brane only in worlvolume orientation so we can apply a suitable parity transformation to turn this into a D5-brane.

Under the $S$ transformation, 
the brane configuration for the $U(N)$ ADHM theory with $l$ flavors maps to that for its mirror theory, 
that is the ADHM circular quiver gauge theory with brane configuration
\begin{align}
\label{IIB_branesetup_ADHMquiver}
\begin{array}{c|cccccccccc}
     & 0 & 1 & 2 & 3 & 4 & 5 & 6 & 7 & 8 & 9 \\ \hline
N \; \textrm{D3} & \circ & \circ & \circ & & & & \circ & & & \\
\textrm{D5} & \circ & \circ & \circ & \circ & \circ & \circ & & & & \\
l \; \textrm{NS5} & \circ & \circ & \circ & & & & & \circ & \circ & \circ \\ \hline
\textrm{D5$'$} & \circ & \circ & & & & & \circ &\circ & \circ & \circ \\
\widetilde{\textrm{NS5$'$}} & \circ & \circ & & \circ & \circ & \circ & \circ & & & 
\end{array}
\end{align}
To achieve this, noting that we would have had an anti-D5-brane and an anti-$\widetilde{\textrm{D5$'$}}$-brane, we also applied parity transformations to reverse the orientation of $x^1$, $x^2$ and $x^3$.
This mirror theory has the $U(N)^l$ twisted vector multiplet consisting of $\mathcal{N}=2$ vector multiplets $\textrm{VM}_{I}$ and adjoint chiral multiplets $\tilde{\Phi}_{I}$, 
with $I=1,\cdots, l$  
and bifundamental twisted hypermultiplets $(T_{I, I+1},\tilde{T}_{I,I+1})$ between adjacent $U(N)_I$ and $U(N)_{I+1}$ gauge nodes 
as well as a single twisted hypermultiplet $(\tilde{I},\tilde{J})$ in the fundamental representation of one of the gauge nodes which we label $U(N)_1$.
The charges of the field content in $\mathcal{N}=2$ language are summarized as 
\begin{align}
\label{mUN_l_ADHMquiver_charges}
\begin{array}{c|c|c|c|c|c|c}
& U(N)_{1}\times \cdots \times U(N)_{l}, & U(1)_u & U(1)_z & U(1)_t & U(1)_x & U(1)_R \\ \hline
\textrm{VM}_{I=1,\cdots, l} & (\underbrace{{\bf 1},\cdots,{\bf 1}}_{I-1},{\bf Adj},{\bf 1},\cdots,{\bf 1}) & 0 & 0 & 0 & 0 & 0 \\
\tilde{\Phi}_{I=1,\cdots, l} & (\underbrace{{\bf 1},\cdots,{\bf 1}}_{I-1},{\bf Adj},{\bf 1},\cdots,{\bf 1}) & 0 & 0  & 2 & 0 & 1 \\
T_{I, I+1} & (\underbrace{{\bf 1},\cdots,{\bf 1}}_{I-1},{\bf N},{\bf \overline{N}},{\bf 1},\cdots,{\bf 1}) & 0 & -1 & -1 & 0 & \frac{1}{2} \\
\tilde{T}_{I, I+1} & (\underbrace{{\bf 1},\cdots,{\bf 1}}_{I-1},{\bf \overline{N}},{\bf N},{\bf 1},\cdots,{\bf 1}) & 0 & 1 & -1 & 0 & \frac{1}{2} \\
\tilde{I} & ({\bf N},{\bf 1},\cdots, {\bf 1}) & 0 & 0 & -1 & 0 & \frac{1}{2} \\
\tilde{J} & ({\bf \overline{N}},{\bf 1},\cdots, {\bf 1}) & 0 & 0 & -1 & 0 & \frac{1}{2} \\
\hline \\
\eta_{I, I+1} & (\underbrace{{\bf 1},\cdots,{\bf 1}}_{I-1},{\bf det},{\bf det^{-1}},{\bf 1},\cdots,{\bf 1}) & 1 & -1 & 0 & 0 & 0 \\
\Gamma & ({\bf N},{\bf 1},\cdots, {\bf 1}) & 0 & 0 & 0 & 1 & 0
\end{array}
\end{align}
where for later convenience we have listed the charges of 2d Fermi multiplets $\eta_{I, I+1}$ and $\Gamma$. Each Fermi $\eta_{I, I+1}$ also has charge $-1$ under the topological $U(1)_{y_I}$ which we didn't include in the table. To be precise, we impose a constraint on the field strengths $\sum_{I = 1}^l y_I = 0$ and the final topological symmetry is provided by $U(1)_x$.

Note that since this is a circular quiver we identify $U(N)_{l+1}$ with $U(N)_1$ etc. In the special case $l = 1$ there is only a single $U(N)$ gauge node and the bifundamental multiplets $T$ and $\tilde{T}$ become adjoint multiplets while the bi-determinant Fermi multiplet becomes a gauge singlet.

When we further introduce the NS5$'$-brane in the brane configuration for the ADHM theory, it maps to the D5$'$-brane. 
This will provide the mirror ADHM quiver gauge theory with Nahm pole boundary conditions for the twisted vector multiplet and hypermultiplets. 
On the other hand, the $\widetilde{\textrm{D5$'$}}$-brane in the brane configuration for the ADHM theory maps to the $\widetilde{\textrm{NS5$'$}}$-brane. 
It leads to Neumann boundary condition for the twisted vector multiplet and hypermultiplets in the mirror ADHM quiver gauge theory.
As previously stated, we do not directly study Nahm pole boundary conditions, but we note that with the assumption of such dualities, studying Neumann boundary conditions in the mirror theory will describe the properties of the theory with Nahm pole boundary conditions, in particular this will give a prediction for the Nahm pole half-index.

\subsubsection{ADHM quiver theory with $(\mathcal{N},N,N)+\Gamma+\eta$}
In the case where we have the $\widetilde{\textrm{NS5$'$}}$-brane in the mirror quiver theory 
we have Neumann boundary conditions for all the multiplets except for the adjoint chirals $\tilde{\Phi}_I$ 
having Dirichlet boundary conditions, giving the following anomaly
\begin{align}
\label{bdy_UN_1_ADHMquiver_anom_NDNN}
\Acal_{\mathcal{N},N,N} = & \sum_I \Bigg[ \underbrace{N \Tr(s_I^2) - \Tr(s_I)^2 + \frac{N^2}{2}r^2}_{\textrm{VM}_I, \; \Ncal} + \underbrace{\left( N \Tr(s_I^2) - \Tr(s_I)^2 + 2N^2t^2 \right)}_{\tilde{\Phi}_I, \; D}
\nonumber \\
 & - \underbrace{\left( N \Tr(s_I^2) + N\Tr(s_{I+1}^2) -2 \Tr(s_I)\Tr(s_{I+1}) + N^2z^2 + N^2\left( -t - \frac{1}{2}r \right)^2 \right)}_{(T, \tilde{T})_{I, I+1}, \; N} \Bigg]
\nonumber \\
 & - \underbrace{\left( \Tr(s_1^2) + N\left( -t - \frac{1}{2}r \right)^2 \right) }_{(\tilde{I}, \tilde{J}), \; N}
  - \underbrace{2xTr(s_1)}_{FI}
  \nonumber \\
  = & -\Tr(s_1^2) - 2\sum_I \Tr(s_I)^2 + 2\sum_I \Tr(s_I)\Tr(s_{I+1}) - 2x\Tr(s_1) \nonumber \\
  & - lN^2 z^2 + (lN-1)Nt^2 - (lN+1)Ntr + \frac{(lN-1)N}{4}r^2 \; .
\end{align}

In addition at each junction of an NS5-brane with the $\widetilde{\textrm{NS5$'$}}$-brane where we also have $N$ D3-branes in two of the quadrants 
we get a bideterminant Fermi $\eta_{I, I+1}$, charged under $U(1)_u$ and $U(1)_z$ \cite{Hanany:2018hlz}, 
giving contribution $(\Tr(s_I) - \Tr(s_{I+1}) - y_I + u - z)^2$, and from the junction of the D5-brane with the $\widetilde{\textrm{NS5$'$}}$-brane 
we get a fundamental Fermi $\Gamma$, charged under the topological $U(1)_x$, giving contribution $\Tr(s_1^2) + 2x \Tr(s_1) + Nx^2$. 
Adding these contributions to the anomaly we have in total
\begin{align}
\label{bdy_UN_l_ADHMquiver_anom_NDNN_Fermi}
\Acal_{\mathcal{N},N,N + \Gamma + \eta} = & lu^2 - 2luz + Nx^2 - l(N^2-1) z^2 + (lN-1)Nt^2
\nonumber\\
& + \sum_{I = 1}^l y_I^2 - (lN+1)Ntr + \frac{(lN-1)N}{4}r^2 \; ,
\end{align}
where we have introduced FI terms $2(\Tr(s_I) - \Tr(s_{I+1}) y_I$ which are cancelled by the $\eta_{I, I+1}$ Fermi contributions.
This is a quantum mechanically consistent $\mathcal{N}=(0,4)$ boundary condition that is free from gauge anomaly, 
which we call $(\mathcal{N},N,N)+\Gamma+\eta$.

In the Abelian case, $N=1$, we have
\begin{align}
\label{bdy_U1_l_ADHMquiver_anom_NDNN_Fermi}
\Acal = & lu^2 - 2luz + x^2 + \sum_{I = 1}^l y_I^2 + (l-1)t^2 - (l+1)tr + \frac{(l-1)}{4}r^2 \; .
\end{align}

\begin{comment}
{\color{red} TODO: Delete or move to Section 4?

which should be compared to the Abelian ADHM theory with Dirichlet boundary conditions which has anomaly \eqref{bdy_UN_l_ADHM_anom_DDD}
\begin{align}
\label{bdy_U1_l_ADHM_anom_DDD}
\Acal = & lu^2 -2luz + x^2 + \Tr(y^2) + (l-1)t^2 - (l+1)tr + \frac{(l-1)}{4}r^2 \; .
\end{align}

These match if we ignore the flavor symmetry contribution $\Tr(y^2)$ 
since matching this is non-trivial as it is a classical symmetry on one side and a quantum symmetry on the other.

Similarly in the case $l = 1$ we have
\begin{align}
\label{bdy_UN_1_ADHMquiver_anom_NDNN_FI_Fermi}
\Acal = & u^2 - 2uz + Nx^2 - (N^2-1) z^2 + (N-1)Nt^2 - (N+1)Ntr + \frac{(N-1)N}{4}r^2 \; .
\end{align}
which matches \eqref{bdy_UN_1_ADHM_anom_NDNN_Gamma} after the mirror mapping $x \leftrightarrow z$ and $t \to -t$.
}
\end{comment}

\subsubsection{ADHM quiver theory with $(\mathcal{D},D,D)$}
In the case where we have the D5$'$-brane in the mirror quiver theory 
we have Dirichlet boundary conditions for all the multiplets except for the adjoint chirals $\tilde{\Phi}_I$ having Neumann boundary conditions, giving the following anomaly
\begin{align}
\label{bdy_UN_1_ADHMquiver_anom_DNDD}
\Acal_{\mathcal{D},D,D} = & \sum_I \Bigg[ \underbrace{-N \Tr(u_I^2) + \Tr(u_I)^2 - \frac{N^2}{2}r^2}_{\textrm{VM}_I, \; \Dcal} - \underbrace{\left( N \Tr(u_I^2) - \Tr(u_I)^2 + 2N^2t^2 \right)}_{\tilde{\Phi}_I, \; N}
\nonumber \\
 & + \underbrace{\left( N \Tr(u_I^2) + N\Tr(u_{I+1}^2) -2 \Tr(u_I)\Tr(u_{I+1}) + N^2z^2 + N^2\left( -t - \frac{1}{2}r \right)^2 \right)}_{(T, \tilde{T})_{I, I+1}, \; D} \Bigg]
\nonumber \\
 & + \underbrace{\left( \Tr(u_1^2) + N\left( -t - \frac{1}{2}r \right)^2 \right) }_{(\tilde{I}, \tilde{J}), \; D}
 - \underbrace{2x\Tr(u_1)}_{FI}
 + \underbrace{2 \sum_I \left( \Tr(u_I) - \Tr(u_{I+1}) \right)y_I}_{FI}
  \nonumber \\
  = & \Tr(u_1^2) - 2x\Tr(u_1) + \sum_I \left( \Tr(u_I) - \Tr(u_{I+1}) \right)^2 + 2 \sum_I \left( \Tr(u_I) - \Tr(u_{I+1}) \right)y_I \nonumber \\
  & + lN^2 z^2 - (lN-1)Nt^2 + (lN+1)Ntr - \frac{(lN-1)N}{4}r^2 \; .
\end{align}
We call this boundary condition $(\mathcal{D},D,D)$, 
where $\mathcal{D}$ is the $\mathcal{N}=(0,4)$ Dirichlet boundary condition for the vector multiplet, 
the other two $D$'s indicate the $\mathcal{N}=(0,4)$ Dirichlet boundary conditions for $(T, \tilde{T})_{I, I+1}$ and $(\tilde{I}, \tilde{J})$.

\begin{comment}
In addition at each junction of an NS5-brane with the NS5$'$-brane where we also have $N$ D3-branes in two of the quadrants we get a bideterminant Fermi giving contribution $(\Tr(s_I) - \Tr(s_{I+1}) + u)^2$ and from the junction of the D5-brane with the NS5$'$-brane we get a fundamental Fermi giving contribution $\Tr(s_1^2)$. Adding these contributions to the anomaly we have in total
\begin{align}
\label{bdy_UN_1_ADHMmirror_anom_NDNN_Fermi}
\Acal = & lu^2 - lN^2 z^2 + (lN-1)Nt^2 - (lN+1)Ntr + \frac{(lN-1)N}{4}r^2 \; .
\end{align}
\end{comment}

In the Abelian case, $N=1$, we have
\begin{align}
\label{bdy_U1_l_ADHMquiver_anom_DNDD}
\Acal = &  u_1^2 + \sum_I \left( u_I - u_{I+1} \right)^2 - 2xu_1 + 2 \sum_I \left( \Tr(u_I) - \Tr(u_{I+1}) \right)y_I \nonumber \\
  & + lz^2 - (l-1)t^2 + (l+1)tr - \frac{(l-1)}{4}r^2 \; .
\end{align}

\begin{comment}
{\color{red} TODO: Delete or move to Section 4?

which should be compared to the Abelian ADHM theory with Neumann boundary conditions which has anomaly \eqref{bdy_U1_l_ADHM_anom_NDNN_Gamma}. We see that the anomalies match up to the flavor symmetry which we ignore since it is a classical symmetry in the ADHM theory but not in the ADHM quiver theory.
}
\end{comment}

\subsection{ABJM and circular quiver Chern-Simons boundary conditions}
\label{Sec_Anom_ABJM}

The $STS$ transformation is easily seen to map NS5 $\to$ $(-1, 0)$ and D5 $\to$ $(-1, -1)$. 
We can follow this with a transformation of $S^2 = -I$ so that finally we have D5 $\to$ $(1, 1)$ and NS5 $\to$ NS5.
More generally, under $-STS$ a $(p,q)$ 5-brane is mapped to a $(p+q, q)$ 5-brane.
Therefore the ADHM brane configuration \eqref{IIB_branesetup_ADHM} maps to
\begin{align}
\label{IIB_branesetup_CQCS}
\begin{array}{c|cccccccccc}
     & 0 & 1 & 2 & 3 & 4 & 5 & 6 & 7 & 8 & 9 \\ \hline
N \; \textrm{D3} & \circ & \circ & \circ & & & & \circ & & & \\
\textrm{NS5} & \circ & \circ & \circ & \circ & \circ & \circ & & & & \\
l \; (1,1) & \circ & \circ & \circ & & & & & \circ & \circ & \circ \\ \hline
\textrm{NS5}' & \circ & \circ & & & & & \circ &\circ & \circ & \circ \\
\widetilde{(1,1)'} & \circ & \circ & & \circ & \circ & \circ & \circ & & & 
\end{array}
\end{align}

This is a somewhat exotic brane configuration due to the $(1,1)$ 5-branes, including the possible boundary brane. We will present a set of boundary conditions for the field theory which we later show to be consistent with expected dualities, at least for the Abelian theories.
The bulk theories and their dualities have been analyzed in \cite{Jafferis:2008em, Imamura:2008nn, Imamura:2008dt, Okazaki:2019ony, Hayashi:2022ldo, Hayashi:2025guk}. As has been indicated by the dualities relating the brane constructions, the bulk ADHM, ADHM quiver and circular quiver Chern-Simons theories are dual to each other. We will explore these dualities in the presence of a boundary in Section~\ref{sec_Dualities}.

The $N$ D3-branes between consecutive $(1,1)$ 5-branes give rise to an $\Ncal = 4$ $U(N)$ twisted vector multiplet with vanishing Chern-Simons level while those between an NS5-brane and a $(1,1)$ 5-brane give rise to an $\Ncal = 2$ $U(N)$ vector multiplet with Chern-Simons level $\pm 1$. This gives the circular quiver Chern-Simons theory with gauge group $U(N)_1 \times U(N)_0^{l-1} \times U(N)_{-1}$. In addition we have $\Ncal = 4$ twisted hypermultiplets in the bifundamental respresentation of neightboring $U(N)$ groups from fundamental strings crossing the $(1, 1)$ 5-branes, along with a bifundamental hypermultiplet from fundamental strings crossing the NS5-brane. In the special case of $l = 1$ we have the ABJM model \cite{Aharony:2008ug}. The field content in terms of $\Ncal = 2$ supermultiplets is summarized in the following table.

\begin{align}
\label{UN_1_CSQuiver_charges}
\begin{array}{c|c|c|c|c|c|c}
& U(N)_1 \times U(N)^{l-1}_0 \times U(N)_{-1} & U(1)_u & U(1)_x & U(1)_z & U(1)_t & U(1)_R \\ \hline
\textrm{VM}_{I=1,\cdots, l+1} & (\underbrace{{\bf 1},\cdots,{\bf 1}}_{I-1},{\bf Adj},{\bf 1},\cdots,{\bf 1}) & 0 & 0 & 0 & 0 & 0 \\
\Phi_{I=2,\cdots, l} & (\underbrace{{\bf 1},\cdots,{\bf 1}}_{I-1},{\bf Adj},{\bf 1},\cdots,{\bf 1}) & 0 & 0 & 0 & 2 & 1 \\
H & ({\bf N},\underbrace{{\bf 1},\cdots,{\bf 1}}_{l-1},{\bf \overline{N}}) & 0 & 1 & 0 & 1 & \frac{1}{2} \\
\widetilde{H} & ({\bf \overline{N}},\underbrace{{\bf 1},\cdots,{\bf 1}}_{l-1},{\bf N}) & 0 & -1 & 0 & 1 & \frac{1}{2} \\
T_{I, I+1 \; (I=1,\cdots, l)} & (\underbrace{{\bf 1},\cdots,{\bf 1}}_{I-1},{\bf N},{\bf \overline{N}},{\bf 1},\cdots,{\bf 1}) & 0 & 0 & 1 & -1 & \frac{1}{2} \\
\widetilde{T}_{I, I+1 \; (I=1,\cdots, l)} & (\underbrace{{\bf 1},\cdots,{\bf 1}}_{I-1},{\bf \overline{N}},{\bf N},{\bf 1},\cdots,{\bf 1}) & 0 & 0 & -1 & -1 & \frac{1}{2} \\
\hline \\
\eta_{I, I+1 \; (I=1,\cdots, l)} & (\underbrace{{\bf 1},\cdots,{\bf 1}}_{I-1},{\bf det},{\bf det^{-1}},{\bf 1},\cdots,{\bf 1}) & 1 & 0 & -1 & 0 & 0 \\
\eta & ({\bf det}, \underbrace{{\bf 1},\cdots,{\bf 1}}_{l-1} ,{\bf det^{-1}}) & 1 & -1 & 0 & 0 & 0
\end{array}
\end{align}
Similarly to the cases of ADHM and ADHM quiver, the Fermis will also be charged under topological $U(1)$ symmetries which have not been included in the table.


If we consider $\Ncal = (0,4)$ boundary conditions for the (twisted) hypermultiplets 
then the two chirals $(H, \widetilde{H})$ in the hypermultiplet will have the same boundary conditions, and similarly for the two chirals $(T_{I, I+1}, \widetilde{T}_{I, I+1})$ in each twisted hypermultiplet. We further simplify to the case that all the twisted hypermultiplets have the same boundary condition. This gives rise to anomaly contributions
\begin{align}
    -\Acal^H_N = \Acal^H_D = & (s_1 - s_{l+1} + x)^2 + \left( t - \frac{1}{2}r \right)^2
\end{align}
for the hypermultiplet with Neumann or Dirichlet boundary condition, and similarly for the twisted hypermultiplets we have
\begin{align}
    -\Acal^T_N = \Acal^T_D = & \sum_{I = 1}^{l} (s_I - s_{I+1} + z)^2 + l \left( t + \frac{1}{2}r \right)^2
    \nonumber \\
    = & \sum_{I = 1}^{l} (s_I - s_{I+1})^2 + 2s_1 z - 2s_{l+1} z + l z^2 + l \left( t + \frac{1}{2}r \right)^2
\end{align}

We argue that the NS5$'$-brane gives rise to Neumann boundary conditions for the hypermultiplet and Dirichlet boundary conditions for the twisted hypermultiplets. For the hypermultiplet this is consistent with expectations arising from the junction of the NS5-brane and NS5$'$-brane so the claim is that 
in the case of a junction between the $(1,1)$-brane and NS5$'$-brane the twisted hypermultiplet gets Dirichlet boundary condition. 
In the case of the $\widetilde{(1,1)'}$-brane we would instead have Dirichlet boundary conditions for the hypermultiplet and Neumann boundary conditions for the twisted hypermultiplets. Before discussing the boundary conditions for the (twisted) vector multiplets, 
we consider the case of $l = 1$.

If we take the case of $l=1$ this maps the brane configuration giving rise to the $U(N)$ ADHM theory with one flavor 
to that giving rise to the $U(N)_1 \times U(N)_{-1}$ ABJM theory with $\Ncal = 2$ multiplets
\begin{align}
\label{UN_1_ABJM_charges}
\begin{array}{c|c|c|c|c|c|c}
& U(N)_1 & U(N)_{-1} & U(1)_x & U(1)_z & U(1)_t & U(1)_R \\ \hline
\textrm{VM} & {\bf Adj} & {\bf 1} & 0 & 0 & 0 & 0 \\
\widetilde{\textrm{VM}} & {\bf 1} & {\bf Adj} & 0 & 0 & 0 & 0 \\
H & {\bf N} & {\bf \overline{N}} & 1 & 0 & 1 & \frac{1}{2} \\
\widetilde{H} & {\bf \overline{N}} & {\bf N} & -1 & 0 & 1 & \frac{1}{2} \\
T & {\bf N} & {\bf \overline{N}} & 0 & 1 & -1 & \frac{1}{2} \\
\widetilde{T} & {\bf \overline{N}} & {\bf N} & 0 & -1 & -1 & \frac{1}{2}
\end{array}
\end{align}

If we replaced the $(1,1)$ 5-brane with a $(1,k)$ 5-brane we would have instead the $U(1)_k \times U(1)_{-k}$ ABJM model. We discuss some aspects of this more general case, although our main focus later will be on the case of $k = 1$ which arises as a dual of the ADHM theory.

We now focus on the case of Abelian ABJM and consider a range of possible boundary conditions, 
although as noted above only specific combinations of boundary conditions will arise from the brane configurations we have considered.
We work in an $\Ncal = 2$ description so allow $\mathcal{N}=(0,2)$ Neumann or Dirichlet boundary conditions \cite{Okazaki:2013kaa} 
independently for the four $\Ncal = 2$ chiral multiplets $H$, $\tilde{H}$, $T$ and $\tilde{T}$ coming from the $\Ncal = 4$ hypermultiplet and twisted hypermultiplet.
These multiplets give anomaly contribution
\begin{align}
\label{bdy_U1_ABJM_HtH_anom_ND}
\Acal_{\textrm{chirals}} = & \pm \underbrace{\left( s_1 - s_2 + x + t - \frac{1}{2}r \right)^2}_{H}
 \pm \underbrace{\left( -s_1 + s_2 - x + t - \frac{1}{2}r \right)^2}_{\tilde{H}}
\nonumber \\
& \pm \underbrace{\left( s_1 - s_2 + z - t - \frac{1}{2}r \right)^2}_{T}
 \pm \underbrace{\left( -s_1 + s_2 - z - t - \frac{1}{2}r \right)^2}_{\tilde{T}}
\end{align}
where in each case we take the plus sign for Dirichlet boundary condition and the minus sign for Neumann boundary condition.

A $U(1)$ VM gives anomaly contribution $\pm \frac{1}{2}r^2$ with sign depending on the choice of Neumann (plus) or Dirichlet (minus) boundary conditions. So, if we have two $U(1)$ VMs with opposite boundary conditions there is no contribution to the anomaly. Assuming this is the case the only additional contribution to the anomaly is from the Chern-Simons levels giving a total anomaly
\begin{align}
\label{bdy_U1_1_ABJM_anom_ND}
\Acal_{CS} = & k(s_1^2 - s_2^2)
\; .
\end{align}
In the case of both vector multiplets having the same boundary condition, the only difference in the Abelian case is an additional contribution $\pm r^2$.

Note that we can take arbitrary linear combinations of the $U(1)_{\pm k}$ vector multiplets and impose boundary conditions on each of the two resulting $U(1)$ vector multiplets. We start by considering such a case where the linear combinations are the sum and difference of the $U(1)_{\pm k}$ vector multiplets. This case is our main focus and we argue that appropriate $\Ncal = (0,2)$ boundary conditions \cite{Okazaki:2013kaa} for these linear combinations lead to a simple half-index for a theory with a residual $\Zb_k$ gauge group.

We also consider standard $\mathcal{N}=(0,2)$ Neumann or Dirichlet boundary conditions \cite{Okazaki:2013kaa} for the original $U(1)_{\pm k}$ vector multiplets. 
Since we want to focus on cases where we don't need to introduce 2d chirals to cancel gauge anomalies 
(for vector multiplets with Neumann boundary conditions) or to introduce 2d chirals or Fermis to produce a convergent monopole flux sum (for vector multiplets with Dirichlet boundary conditions), there are constraints of the allowed boundary conditions. 
We present the analysis in Appendix~\ref{ABJM_U1k_VMbc}.

\subsubsection{Diagonal gauge group breaking -- $\Zb_k$ gauge theory}
\label{Sec_ABJM_U1k_diag}
We consider the boundary conditions for the vector multiplets 
as in section~\ref{Sec_Anom_ABJM} with a small modification that now we have Chern-Simons levels $k$ and $-k$ where wlog.\ we take $k > 0$. 
Again defining the two linear combinations of the $U(1)_{\pm k}$ field strengths $s_{+} = \frac{1}{2}(s_1 + s_2)$ for the diagonal subgroup, 
$U(1)_{+}$ and $s_{-} = s_1 - s_2$ for the other combination, $U(1)_{-}$, 
we then have anomaly contribution from the vector multiplets and Chern-Simons levels
\begin{align}
\label{bdy_U1_k_ABJM_anom_VMdiag}
\Acal_{CS} = & 2k s_{+} s_{-}
\; .
\end{align}
up to an additional term $-r^2$, $0$ or $r^2$ depending on the boundary conditions for $U(1)_{\pm}$.
Since the chirals are charged under $U(1)_{-}$ but not $U(1)_{+}$ then no matter which boundary conditions we choose for the chirals we have no gauge anomaly for $U(1)_{+}$ but do have a mixed anomaly from the Chern-Simons contributions.

Now, since we have no gauge anomaly for $U(1)_{+}$, we will not get a convergent half-index if we have the standard Dirichlet boundary condition for the $U(1)_{+}$ vector multiplet (at least without introducing 2d Fermis to shift the effective Chern-Simons level). This rules out the possibility of Dirichlet boundary conditions for both vector multiplets.

Now consider Neumann boundary condition for the $U(1)_{+}$ vector multiplet. As we do not have a gauge anomaly for $U(1)_{+}$ we are not required to introduce any 2d chiral or Fermi multiplets and here we choose not to do so. Due to the mixed anomaly this breaks the $U(1)_{-}$ symmetry, although we argue that a discrete $\Zb_k$ gauge group remains. Since the $U(1)_{-}$ gauge group is broken it is consistent to assume that the $U(1)_{-}$ vector multiplet has Dirichlet\footnote{It is not clear how to interpret the case where both vector multiplets have Neumann boundary conditions with such a mixed anomaly term.} boundary conditions.
We are then left with the 4 free chirals $H$, $\tilde{H}$, $T$ and $\tilde{T}$ with arbitrary boundary conditions, along with any 2d matter contributions.
We expect this combination of vector multiplet boundary conditions to arise from the NS5$'$-brane, although we note that in that case we also fix the boundary conditions to be Neumann for the hypermultiplet and Dirichlet for the twisted hypermultiplet.

If instead we take Neumann boundary conditions for the $U(1)_{-}$ vector multiplet then we must deal with the issue that the boundary conditions for the chirals would lead to a gauge anomaly which must be cancelled, except for the case of equal numbers of Neumann and Dirichlet boundary conditions. In all cases we would expect to be left with a mixed anomaly involving the $U(1)_{+}$ together with a linear combination of the $U(1)_x$, $U(1)_z$, $U(1)_t$ and $U(1)_R$ symmetries which would break that linear combination, possibly to a $\Zb_k$ symmetry. We expect this choice of vector multiplet boundary conditions, together with Dirichlet for the hypermultiplet and Neumann for the twisted hypermultiplet, to arise from the $\widetilde{(1,1)'}$ 5-brane.

Now, while the choice of Neumann boundary conditions for the $U(1)_{+}$ vector multiplet and Dirichlet boundary conditions for the $U(1)_{-}$ vector multiplet seems to lead to a complete breaking of the $U(1)_{-}$ gauge group, we argue that there is a residual unbroken $\Zb_k$ gauge group\footnote{A similar situation arises if we have Neumann boundary conditions for the $U(1)_{-}$ vector multiplet and Dirichlet for the $U(1)_{+}$ vector multiplet although we note that it would be a combination of the $U(1)_{+}$ with other global symmetries which is broken to $\Zb_k$.}. This is most straightforwardly justified by considering the bulk theory where the opposite Chern-Simons levels are reexpressed through the linear combinations of the sum and difference of those $U(1)$ gauge fields as a BF term with level $k$ in the action. I.e.\ if we define $B^{\pm} = A^{(1)} \pm A^{(2)}$ where $A^{(1)}$ and $A^{(2)}$ are the $U(1)_k$ and $U(1)_{-k}$ gauge potentials then the Chern-Simons action takes the form
\begin{align}
    \Lcal = & \frac{k}{4\pi} \left( A^{(1)} \wedge dA^{(1)} - A^{(2)} \wedge dA^{(2)} \right)
    = \frac{k}{4\pi} B^{-} \wedge dB^{+} + \frac{k}{8\pi} d(B^{-} \wedge B^{+})
    \nonumber \\
    & = \frac{k}{2\pi} A^{1} \wedge dB^{+} - \frac{k}{4\pi} B^{+} \wedge dB^{+} + \frac{k}{8\pi} d(B^{-} \wedge B^{+})
    \nonumber \\
    & = \frac{k}{4\pi} B^{+} \wedge dB^{-} + \frac{k}{8\pi} d(B^{+} \wedge B^{-})
    \nonumber \\
    & = \frac{k}{2\pi} A^{2} \wedge dB^{-} + \frac{k}{4\pi} B^{-} \wedge dB^{-} + \frac{k}{8\pi} d(B^{+} \wedge B^{-})
    \; .
\end{align}
In the bulk theory without boundary the total derivative term $d(B^{-} \wedge B^{+})$ can be ignored. When we have a boundary this term is also trivial in the case where we have boundary condition either $B^{+} = 0$ or $B^{-} = 0$ which is exactly what we argued for. This then leaves a standard BF theory.
Such a theory with BF term of the form $\frac{k}{2 \pi} B \wedge dA$ is known \cite{Banks:2010zn,Kapustin:2014gua} to be equivalent to a $\Zb_k$ gauge theory 
and so we claim that the resulting system keeps residual $\Zb_k$ gauge group on the boundary which couples to the 3d chiral multiplets. 
Note that we have a $\Zb_k$ gauge theory from the expressions containing $\frac{k}{2\pi} A^{1} \wedge dB^{+}$ or $\frac{k}{2\pi} A^{2} \wedge dB^{-}$. 
The expressions with $\frac{k}{4\pi} B^{-} \wedge dB^{+}$ or $\frac{k}{4\pi} B^{+} \wedge dB^{-}$ appear equivalent but with the wrong normalization. 
The explanation is that the gauge fields must obey the standard normalization condition that for any closed 2-cycle $\Sigma$, $\frac{1}{2\pi} \int_{\Sigma} A \in \Zb$. 
If this is the case for $A^{(1)}$ and $A^{(2)}$ then obviously it is also true for $B^{\pm}$. 
However, for $B^{\pm}$ the integers are not totally independent. 
In particular both are either even or odd whereas there is no such restriction on the pairs $(A^{(1)}, B^{+})$ or $(A^{(2)}, B^{-})$.


We also assume that we have the bi-determinant Fermi, which in the Abelian case considered here has $U(1)_k \times U(1)_{-k}$ charges $(+1, -1)$ as well as charge $+1$ under the global $U(1)_u$ and $-1$ under the global $U(1)_x$. The charge under the gauge group is then charge $+1$ under $U(1)_{-}$ and uncharged under $U(1)_{+}$. This reduces to being charged under the residual $\Zb_k$ gauge group only.

Note that in the case of $k=1$ we have just free chiral multiplets with no discrete gauge group.

For the circular quiver theory with $l > 1$ 
we have a similar set of boundary conditions for the linear combinations of the $U(1)_{\pm 1}$ vector multiplets 
(and we could also consider generalizing to Chern-Simons levels $\pm k$). 
In particular if we have the NS5$'$-brane then we have Neumann boundary condition for the $U(1)_{+}$ vector multiplet and Dirichlet for the $U(1)_{-}$ vector multiplet, 
along with Neumann for the hypermultiplet and Dirichlet for the twisted hypermultiplets. 
The (twisted) hypermultiplets then give anomaly contribution
\begin{align}
    \Acal^{H, T} = & \sum_{I = 1}^{l} (s_I - s_{I+1})^2 - (s_1 - s_{l+1})^2 + 2(s_1 - s_{l+1})(z - x) + l z^2 
    \nonumber\\
    &+ l \left( t + \frac{1}{2}r \right)^2 - \left( t - \frac{1}{2}r \right)^2 - x^2
\end{align}
and we should also include the bideterminant Fermi $\eta$ from the intersection of the NS5-brane with the NS5$'$-brane, so we end up with
\begin{align}
    \Acal = & \sum_{I = 1}^{l} (s_I - s_{I+1})^2 + 2(s_1 - s_{l+1})(z + u - x) + l z^2 + u^2 
    \nonumber\\
    &- 2ux + (l-1)t^2 +(l+1)tr + \frac{l-1}{4}r^2 \; .
\end{align}
We claim that the NS5$'$-brane results in Dirichlet boundary conditions for the twisted vector multiplets, 
which in $\Ncal = 2$ notation means $\mathcal{N}=(0,2)$ Dirichlet for the vector multiplets 
and $\mathcal{N}=(0,2)$ Neumann for the adjoint chiral multiplets \cite{Okazaki:2013kaa}, 
so we need to add contribution
\begin{align}
    & -2 (l-1)t^2 - \frac{l-1}{2}r^2 \; .
\end{align}
Including the Chern-Simons contributions we have the total anomaly
\begin{align}
    \Acal = & \sum_{I = 1}^{l} (s_I - s_{I+1})^2 + (s_1 + s_{l+1})(s_1 - s_{l+1}) + 2(s_1 - s_{l+1})(z + u - x) \nonumber \\
    & + l z^2 + u^2 - 2ux - (l-1)t^2 +(l+1)tr - \frac{l-1}{4}r^2 \; .
\end{align}
More precisely, we interpret the vector multiplet boundary conditions as Neumann for $U(1)_{+}$ with field strength $s_{+} = \frac{1}{2}(s_1 + s_{l+1})$ and Dirichlet for the linear combinations with field strengths $v_I = s_I - s_{I+1}$. Note that by the cyclic property we have $\sum_{I = 1}^l v_I = s_1 - s_{l+1} = -v_{l+1}$. In this notation the anomaly is
\begin{align}
    \Acal = & \sum_{I = 1}^{l} v_I^2 + 2(s_{+} + z + u - x) \sum_{I = 1}^l v_I  \nonumber \\
    & + l z^2 + u^2 - 2ux - (l-1)t^2 +(l+1)tr - \frac{l-1}{4}r^2 \; .
\end{align}

If we take into account the total breaking of the $U(1)_{-}$ gauge group we should set $-v_{l+1} = s_1 - s_{l+1} = 0$ and the anomaly simplifies to
\begin{align}
\label{bdy_U1_lp1_CQCS_anom_DND_eta_origvar}
    \Acal_{\Dcal, N, D + \eta} = & \sum_{I = 1}^{l} v_I^2 + l z^2 + u^2 - 2ux - (l-1)t^2 +(l+1)tr - \frac{l-1}{4}r^2
\end{align}
where now $\sum_{I = 1}^l v_I = 0$. This constraint means that the $U(1)_{v_I}$ provide a global $U(1)^{l-1}$ symmetry, with a further $U(1)_u$ symmetry. So, we could change variables to $v_I = u_I - u_{I+1}$ with $u_{l+1} \equiv u_1$ and we are also free to label $u \equiv u_1$ giving the anomaly in a form convenient for matching to dual theories
\begin{align}
\label{bdy_U1_lp1_CQCS_anom_DND_eta}
    \Acal_{\Dcal, N, D + \eta} = & \sum_{I = 1}^{l} (u_I - u_{I+1})^2 + l z^2 + u_1^2 - 2u_1x - (l-1)t^2 +(l+1)tr - \frac{l-1}{4}r^2
\end{align}

Note that after the diagonal breaking and decoupling of $U(1)_{+}$ we have field content and charges
\begin{align}
\label{U1_l_CSQuiver_charges_reduced}
\begin{array}{c|c|c|c|c|c|c}
& U(1)_{v_I}^{l-1} & U(1)_u & U(1)_x & U(1)_z & U(1)_t & U(1)_R \\ \hline
\textrm{VM}_{I=2,\cdots, l} & {\bf 0} & 0 & 0 & 0 & 0 & 0 \\
\Phi_{I=2,\cdots, l} & {\bf 0} & 0 & 0 & 0 & 2 & 1 \\
H & {\bf 0} & 0 & 1 & 0 & 1 & \frac{1}{2} \\
\widetilde{H} & {\bf 0} & 0 & -1 & 0 & 1 & \frac{1}{2} \\
T_l & (-1,\cdots,-1) & 0 & 0 & 1 & -1 & \frac{1}{2} \\
\widetilde{T}_l & (1,\cdots,1) & 0 & 0 & -1 & -1 & \frac{1}{2} \\
T_{I \; (I=1,\cdots, l-1)} & (\underbrace{0,\cdots,0}_{I-1},1,0,\cdots,0) & 0 & 0 & 1 & -1 & \frac{1}{2} \\
\widetilde{T}_{I \; (I=1,\cdots, l-1)} & (\underbrace{0,\cdots,0}_{I-1},-1,0,\cdots,0) & 0 & 0 & -1 & -1 & \frac{1}{2} \\
\hline \\
\eta_l & (-1,\cdots,-1) & 1 & 0 & -1 & 0 & 0 \\
\eta_{I \; (I=1,\cdots, l-1)} & (\underbrace{0,\cdots,0}_{I-1},1,0,\cdots,0) & 1 & 0 & -1 & 0 & 0 \\
\eta & {\bf 0} & 1 & -1 & 0 & 0 & 0
\end{array}
\end{align}


If instead of the NS5$'$-brane we choose the $\widetilde{(1,1)'}$ 5-brane as the boundary brane we claim that we get the same diagonal breaking of $U(1)_1 \times U(1)_{-1}$ but that the other boundary conditions are the opposite ones. In addition, we no longer have the Fermi $\eta$ but instead from the intersection of the $l$ $(1,1)$ 5-branes with the $\widetilde{(1,1)'}$ 5-brane we get $l$ Fermis $\eta_I$. These Fermis exactly cancel the $U(1)^{l-1}$ gauge anomaly and, defining $s_l = - \sum_{I = 1}^{l-1} s_I$, the final result is anomaly
\begin{align}
\label{bdy_U1_lp1_CQCS_anom_NDN_eta_l}
\Acal_{\Ncal, D, N + \eta_I} = &
\underbrace{\frac{(l-1)}{2}r^2}_{\textrm{VM}_I, \; \Ncal}
+ \underbrace{2(l-1)t^2}_{\tilde{\Phi}_I, \; D}
- \sum_{I=1}^{l} \underbrace{\left( s_I^2 + 2s_I z + z^2 +\left( -t - \frac{1}{2}r \right)^2 \right)}_{(T, \tilde{T})_{I}, \; N}
\nonumber \\
 & + \underbrace{\left( x^2 + \left( t - \frac{1}{2}r \right)^2 \right) }_{(H, \widetilde{H}), \; D}
  + \sum_{I=1}^{l} \underbrace{\left( s_I + u - z \right)^2}_{\eta_I}
  \nonumber \\
  = & l u^2 - 2luz + x^2 + (l-1)t^2 - (l+1)tr + \frac{l-1}{4}r^2 \; .
\end{align}
It is not completely clear why the diagonal breaking happens in exactly the same way, but we will later see support for this through dualities checked by matching of anomalies and half-indices.

\section{Half-indices}
\label{sec_hindex}
Half-indices are supersymmetric indices for theories with a boundary. 
Since Neumann or Dirichlet boundary conditions project out half the components of each supermultiplet, 
the contributions of a multiplet to the indices essentially factorize as a product of the Dirichlet and Neumann half-index contributions 
\cite{Dimofte:2017tpi} (see also \cite{Gadde:2013wq,Gadde:2013sca,Yoshida:2014ssa}). 

In the computation of the half-indices, it is convenient to introduce the $q$-shifted factorial defined by
\begin{align}
\label{qpoch_def}
(a;q)_{0}&:=1,\qquad
(a;q)_{n}:=\prod_{k=0}^{n-1}(1-aq^{k}),\qquad 
(q)_{n}:=\prod_{k=1}^{n}(1-q^{k}),
\nonumber \\
(a;q)_{\infty}&:=\prod_{k=0}^{\infty}(1-aq^{k}),\qquad 
(q)_{\infty}:=\prod_{k=1}^{\infty} (1-q^k), 
\end{align}
where $a$ and $q$ are complex numbers with $|q|<1$. 
For simplicity we use the following notation: 
\begin{align}
(x^{\pm};q)_{n}:=(x;q)_{n}(x^{-1};q)_{n}. 
\end{align}

For the vector multiplet we also have either an integral over the gauge fugacities for Neumann boundary condition or a sum over monopole fluxes for Dirichlet boundary conditions. 

Tthe $\mathcal{N}=2$ vector multiplet half-index contribution for $U(N)$ theories is
\begin{align}
    \frac{(q)_{\infty}^{N}}{N!} \oint \left( \prod_{i = 1}^{N} \frac{ds_i}{2\pi i s_i} \right) \prod_{1 \le i < j \le N} (s_i^{\pm} s_j^{\mp}; q)_{\infty}, 
\end{align}
where the integral ensures the gauge invariance.

If, instead, we have Dirichlet boundary condition for the $\mathcal{N}=2$ vector multiplet the contribution is 
\begin{align}
    \frac{1}{(q)_{\infty}^N} \sum_{m_i \in \Zb} \frac{(-q^{1/2})^{k_{eff}[m, m]} u^{k_{eff}[m, -]}}{\prod_{i, j = 1}^N (q^{1 + m_i - m_j} u_i u_j^{-1}; 1)_{\infty}}, 
\end{align}
where expressions with $k_{eff}$ are determined by the anomaly polynomial and we use $u_i$ for the gauge fugacities in the case of Dirichlet boundary conditions. In particular, if the only gauge term appearing in the anomaly polynomial is $\tilde{N} \Tr(u^2)$ then
\begin{align}
    k_{eff}[m, m] & = \tilde{N} \sum_{i = 1}^N m_i^2 \; , \\
    u^{k_{eff}[m, -]} & = \prod_{i = 1}^N u_i^{\tilde{N} m_i} \; .
\end{align}
In addition to this contribution, the contribution for all matter multiplets is modified from the expressions below by the replacement $s_i \to q^{m_i} u_i$.

Including $N_f$ fundamental 3d chirals with R-charge $r$ and Neumann boundary conditions we should include a factor 
\begin{align}
    \prod_{\alpha = 1}^{N_f} \prod_{i = 1}^N \frac{1}{(q^{\frac{r}{2}} s_i x_{\alpha}; q)_{\infty}}, 
\end{align}
where $x_{\alpha}$ are $U(N_f)$ flavor symmetry fugacities (which are often split into a $U(1)$ axial and $SU(N_f)$ flavor symmetry fugacities). Anti-fundamental chirals give a similar contribution with $s_i \rightarrow s_i^{-1}$ and $x_{\alpha} \rightarrow \tilde{x}_I$ for the $U(N_a)$ global symmetry.

Instead, Dirichlet boundary conditions for a fundamental would give contribution
\begin{align}
    \prod_{\alpha = 1}^{N_f} \prod_{i = 1}^N (q^{1 - \frac{r}{2}} s_i^{-1} x_{\alpha}^{-1}; q)_{\infty} \; .
\end{align}

Similarly, for an adjoint chiral with Neumann boundary condition the contribution is
\begin{align}
    \prod_{i, j = 1}^N \frac{1}{(q^{\frac{r}{2}} s_i s_j^{-1} x; q)_{\infty}}, 
\end{align}
where $x$ is a flavor fugacity.
For Dirichlet boundary condition the contribution is
\begin{align}
    \prod_{i, j = 1}^N (q^{1 - \frac{r}{2}} s_i s_j^{-1} x^{-1}; q)_{\infty} \; .
\end{align}

On the boundary we can have 2d Fermi or chiral multiplets. In the fundamental representation these Fermi multiplets with R-charge $0$ give the same contribution as the combination of a pair of fundamental and anti-fundamental 3d chirals with R-charge $r = 1$ and Dirichlet boundary conditions up to an identification of flavor fugacities. In particular, for $M$ fundamental 2d Fermi multiplets we have contribution
\begin{align}
    \prod_{\alpha = 1}^{M} \prod_{i = 1}^N (q^{\frac{1}{2}} s_i^{\pm} x_{\alpha}^{\pm}; q)_{\infty} \; .
\end{align}
where the Fermis are in the fundamental representation of a global $U(M)$ symmetry. 

To make the above statement precise, this is equivalent to the case of $N_f = N_a = M$ where the 3d chirals have Dirichlet boundary conditions and we have specialized the $U(N_a)$ flavor fugacities $\tilde{x}_{\alpha} \rightarrow x_{\alpha}^{-1}$.

Also, note that with further identification of flavor fugacities, a 2d fundamental Fermi together with a fundamental 3d chiral with Neumann boundary conditions and R-charge $1$ gives the same contribution as a 3d fundamental chiral with Dirichlet boundary conditions and R-charge $1$. Specifically, there is a partial cancellation in the half-index contribution if the 3d chiral and 2d Fermi both have the same flavor fugacity $x_{\alpha}$.

\subsection{$U(N)$ ADHM theory}
We present the formulae of the half-indices for the basic $\mathcal{N}=(0,4)$ supersymmetric boundary conditions in the $U(N)$ ADHM theory. 

\subsubsection{ADHM theory with $(\mathcal{N},N,N)+\widetilde{\Gamma}+\widetilde{\eta}$}
We begin with the $\mathcal{N}=(0,4)$ boundary condition $(\mathcal{N},N,N)+\widetilde{\Gamma}+\widetilde{\eta}$ 
in the $U(N)$ ADHM theory with $l$ flavors. 
The half-index can be evaluated from the matrix integral of the form
\begin{align}
\label{ADHMuN_l_04Neu}
&
\mathbb{II}_{\mathcal{N},N,N+\widetilde{\Gamma}+\widetilde{\eta}}^{\textrm{$U(N)$ ADHM}-[l]}(t,x,z,u_I, y_I;q)
\nonumber\\
&=\frac{1}{N!} \frac{(q)_{\infty}^N (q^{\frac12}t^2;q)_{\infty}^N (q^{\frac{1}{2}} u_1^{\pm} x^{\mp}; q)_{\infty}}{(q^{\frac14}tx^{\pm};q)_{\infty}^N}
\oint \prod_{i=1}^{N}\frac{ds_i}{2\pi is_i}
\prod_{i<j}
(s_i^{\pm}s_j^{\mp};q)_{\infty}
(q^{\frac12}t^2s_i^{\pm}s_j^{\mp};q)_{\infty}
\nonumber\\
&\times \left( \prod_{i<j}
\frac{1}{(q^{\frac14}ts_i^{\pm}s_j^{\mp}x^{\pm};q)_{\infty}(q^{\frac14}ts_i^{\pm}s_j^{\mp}x^{\mp};q)_{\infty}} \right)
\prod_{i=1}^N \prod_{I = 1}^l
\frac{(q^{\frac12}s_i^{\pm} u_I^{\pm} u_{I+1}^{\mp} y_I^{\pm} z^{\mp};q)_{\infty}}
{(q^{\frac14}ts_i^{\pm} y_I^{\pm};q)_{\infty}}, 
\end{align}
where the fugacities can be identified from \eqref{UN_l_ADHM_charges} and note that $\prod_{I = 1}^l y_I = 1$.
Note that the half-index (\ref{ADHMuN_l_04Neu}) is also well-defined for 3d $\mathcal{N}=8$ $U(N)$ SYM theory with $l=0$ though the full-index is not.

In the Abelian case $N = 1$ the half-index \eqref{ADHMuN_l_04Neu} becomes 
\begin{align}
\label{ADHMu1_l_04Neu}
\mathbb{II}_{\mathcal{N},N,N+\widetilde{\Gamma}+\widetilde{\eta}}^{\textrm{$U(1)$ ADHM}-[l]}
&=(q)_{\infty}(q^{\frac12}t^2;q)_{\infty}
\frac{(q^{\frac{1}{2}} u_1^{\pm} x^{\mp}; q)_{\infty}}{(q^{\frac14}tx^{\pm};q)_{\infty}}
\oint \frac{ds}{2\pi is} \prod_{I = 1}^l
\frac{(q^{\frac12}s^{\pm} u_I^{\pm} u_{I+1}^{\mp} y_I^{\pm} z^{\mp};q)_{\infty}}
{(q^{\frac14}ts^{\pm} y_I^{\pm};q)_{\infty}}. 
\end{align}
while setting also $l = 1$ the half-index becomes
\begin{align}
\label{ADHMu1_1_04Neu}
\mathbb{II}_{\mathcal{N},N,N+\widetilde{\Gamma}+\widetilde{\eta}}^{\textrm{$U(1)$ ADHM}-[1]}
&=(q)_{\infty}(q^{\frac12}t^2;q)_{\infty}
\frac{(q^{\frac{1}{2}} u_1^{\pm} x^{\mp}; q)_{\infty}}{(q^{\frac14}tx^{\pm};q)_{\infty}}
\oint \frac{ds}{2\pi is}
\frac{(q^{\frac12}s^{\pm}z^{\mp};q)_{\infty}}
{(q^{\frac14}ts^{\pm};q)_{\infty}}. 
\end{align}

We remark on several special fugacity limits of the half-index (\ref{ADHMuN_l_04Neu}). 
Taking the H-twist limit $t\rightarrow q^{1/4}$ \cite{Gaiotto:2019jvo,Okazaki:2019bok}, turning off the flavor fugacities $y_{\alpha}$, $u_I$, $x$ and $z$, 
the integral (\ref{ADHMuN_l_04Neu}) becomes independent of $l$ and it reduces to 
\begin{align}
&
\mathbb{II}_{\mathcal{N},N,N+\widetilde{\Gamma}+\widetilde{\eta}}^{{\textrm{$U(N)$ ADHM}-[l]}^{(H)}}(q)
\nonumber\\
&=\frac{1}{N!}\frac{(q)_{\infty}^{2N}}{(q^{\frac12};q)_{\infty}^{2N - 2}}
\oint \prod_{i=1}^{N}\frac{ds_i}{2\pi is_i}
\frac{\prod_{i<j} (s_i^{\pm}s_j^{\mp};q)_{\infty} (qs_i^{\pm}s_j^{\mp};q)_{\infty}}
{\prod_{i<j} (q^{\frac12}s_i^{\pm}s_j^{\mp};q)_{\infty}^2}. 
\end{align}
If we divide this expression by $(q^{\frac12};q)_{\infty}^2$, i.e.\ if we had not included the contribution from the neutral Fermi $\tilde{\eta}$, this is the unflavored Schur index of $\mathcal{N}=4$ $U(N)$ SYM theory, 
for which several exact closed-form expressions are known \cite{Bourdier:2015wda,Pan:2021mrw,Beem:2021zvt,Huang:2022bry,Hatsuda:2022xdv}. 

When we take the Higgs limit \cite{Razamat:2014pta} where $\mathfrak{t}$ $:=$ $q^{1/4}t$ is kept finite and $q$ is set to $0$, 
the half-index (\ref{ADHMuN_l_04Neu}) becomes 
\begin{align}
&
\mathbb{II}_{\mathcal{N},N,N+\widetilde{\Gamma}+\widetilde{\eta}, \textrm{Higgs}}^{\textrm{$U(N)$ ADHM}-[l]}(x,y_{\alpha};\mathfrak{t})
\nonumber\\
&=\frac{1}{N!}\frac{(1-\mathfrak{t}^2)^N}{(1-\mathfrak{t}x^{\pm})^N}
\oint \prod_{i=1}^N \frac{ds_i}{2\pi is_i}
\frac{\prod_{i\neq j}\left(1-\frac{s_i}{s_j}\right)\left(1-\mathfrak{t}^2\frac{s_i}{s_j}\right)}
{\prod_{i\neq j} \left(1-\mathfrak{t}\frac{s_i}{s_j}x^{\pm}\right)}
\prod_{i=1}^N\prod_{\alpha=1}^{l} \frac{1}{(1-\mathfrak{t}s_i^{\pm}y_{\alpha}^{\pm})}, 
\end{align}
which is identified with the Higgs index of the $U(N)$ ADHM theory, 
i.e.\ the refined Hilbert series for the moduli space of $N$ $SU(l)$ instantons \cite{Benvenuti:2010pq,Hanany:2012dm}. 
Although the closed-form expression for general $N$ and $l$ is not known, 
the cases with special values for $N$ and $l$ can be computed by picking up the JK residues, 
by making use of the refined topological vertex \cite{Iqbal:2003ix,Iqbal:2003zz,Eguchi:2003sj,Taki:2007dh,Awata:2008ed} 
based on the relation to the Nekrasov partition function of 5d $\mathcal{N}=1$ $SU(l)$ Yang-Mills theory with instanton number $N$ 
or by expanding the integrand in terms of the Hall-Littlewood functions \cite{Crew:2020psc,Hayashi:2024jof}. 

While the half-index (\ref{ADHMuN_l_04Neu}) diverges in the large $l$ limit, it possesses a large $N$ limit expression. 
By employing the techniques of the matrix models (see e.g. \cite{Aharony:2003sx,Kinney:2005ej}) for $l=0$, i.e.\ $\mathcal{N}=8$ $U(N)$ SYM theory, 
we obtain the large $N$ half-index
\begin{align}
\label{largeN_N=8}
\mathbb{II}_{\mathcal{N},N,N+\widetilde{\Gamma}+\widetilde{\eta}}^{\textrm{$\mathcal{N}=8$ $U(\infty)$ SYM}}(t,x,u_1;q)
&=\prod_{n=1}^{\infty}
\frac{(1-q^n)(1-u_1x^{-1}q^{n-1/2})(1-u_1^{-1}xq^{n-1/2})}
{(1-q^{\frac{n}{4}}t^n x^n) (1-q^{\frac{n}{4}}t^nx^{-n})}. 
\end{align}
Taking the plethystic logarithm of the expression (\ref{largeN_N=8}), we get the single particle half-index 
\begin{align}
ii_{\mathcal{N},N,N+\widetilde{\Gamma}+\widetilde{\eta}}^{\textrm{$\mathcal{N}=8$ $U(\infty)$ SYM}}(t,x,u_1;q)
&=-\frac{q}{1-q}-\frac{q^{\frac12}u_1x^{-1}}{1-q}-\frac{q^{\frac12}u_1x^{-1}}{1-q}
\nonumber\\
&+\frac{q^{\frac14}tx}{1-q^{\frac14}tx}+\frac{q^{\frac14}tx^{-1}}{1-q^{\frac14}tx^{-1}}. 
\end{align}
Furthermore, setting $t$, $x$, $u_{I}$, $z$ and $y_{\alpha}$ to unity, 
we find that the large $N$ limit of the half-index (\ref{ADHMuN_l_04Neu}) for the ADHM theory with $l\ge 1$ is given by
\begin{align}
\label{ADHMlarge_04Neu}
\mathbb{II}_{\mathcal{N},N,N+ \eta + \Gamma}^{\textrm{$U(\infty)$ ADHM}-[l]}(q)
&=\prod_{n=1}^{\infty}\frac{1}{(1-q^{n-\frac34})^2(1-q^{n-\frac12})^{l^2}(1-q^{n-\frac14})^2(1-q^n)}. 
\end{align}
The single particle half-index is
\begin{align}
\label{ADHMlarge_04Neu_single}
ii_{\mathcal{N},N,N+ \eta + \Gamma}^{\textrm{$U(\infty)$ ADHM}-[l]}(q)
&=\frac{2q^{\frac14}+l^2 q^{\frac12}+2q^{\frac34}+q}{1-q}. 
\end{align}
When one expands the large $N$ half-index (\ref{ADHMlarge_04Neu}) as
\begin{align}
\mathbb{II}_{\mathcal{N},N,N+ \eta + \Gamma}^{\textrm{$U(\infty)$ ADHM}-[l]}(q)
&=\sum_{n\ge 0}d(n) q^{\frac{n}{4}}, 
\end{align}
the expansion coefficient $d(n)$ is identified with the degeneracy of the boundary BPS local operators. 
As $n\rightarrow \infty$, the asymptotic growth of the degeneracy can be evaluated 
by performing the convolution of the asymptotic coefficients for the infinite products in (\ref{ADHMlarge_04Neu}) 
obtained from the Meinardus Theorem \cite{MR62781,MR1634067}. 
We find 
\begin{align}
\label{asymp_dn}
d(n)&\sim \frac{\sqrt{l^2+5}}{2^{\frac{l^2}{2}+3} 3^{\frac12}n}
\exp\left[
\pi\sqrt{\frac{l^2+5}{6}} n^{\frac12}
\right]. 
\end{align}
For example, for $l=1$, $2$ and $3$, 
the exact numbers $d(n)$ and the values $d_{\textrm{asymp}}(n)$ evaluated from the formula (\ref{asymp_dn}) are shown as follows:
\begin{align}
l=1: \quad 
\begin{array}{c|c|c}
n&d(n)&d_{\textrm{asymp}}(n) \\ \hline
10&232&257.897\\
100&5.32874\times 10^{10}&5.50394\times 10^{10}\\
1000&1.72936\times 10^{39}&1.74694\times 10^{39}\\
\end{array}, \\
l=2: \quad 
\begin{array}{c|c|c}
n&d(n)&d_{\textrm{asymp}}(n) \\ \hline
10&1092&1041.4\\
100&2.83079\times 10^{13}&2.77677\times 10^{13}\\
1000&3.78785\times 10^{48}&3.76253\times 10^{48}\\
\end{array}, \\
l=3: \quad 
\begin{array}{c|c|c}
n&d(n)&d_{\textrm{asymp}}(n) \\ \hline
10&6928&4648.82\\
100&9.44781\times 10^{16}&8.27888\times 10^{16}\\
1000&1.00282\times 10^{61}&9.60324\times 10^{60}\\
\end{array}. 
\end{align}
The large $N$ single particle half-index (\ref{ADHMlarge_04Neu_single}) is expected 
to encode the spectrum of the Kaluza-Klein (KK) modes of the massless fields in the holographically dual supergravity background. 
The finite $N$ correction that shows up at order $q^{\frac{N+1}{4}}$ will be contributed from the giant gravitons in the dual geometry.  
We hope to pursue the gravity dual analysis further elsewhere.

\subsubsection{ADHM theory with $(\mathcal{D},D,D)$}
Next consider the $\mathcal{N}=(0,4)$ boundary condition $(\mathcal{D},D,D)$ for the $U(N)$ ADHM theory with $l$ flavors. 
In this case, the gauge group is broken down to the boundary global symmetry. 
The half-index is given by
\begin{align}
\label{h_ADHMuNnfl_DDD}
&
\mathbb{II}_{\mathcal{D},D,D}^{\textrm{$U(N)$ ADHM}-[l]}(t,x,z,u_i,y_{\alpha};q)
\nonumber\\
& = \frac{(q^{\frac{3}{4}} t^{-1} x^{\pm}; q)_{\infty}^N}{(q)_{\infty}^N (q^{\frac12}t^{-2};q)_{\infty}^N} \sum_{m_i \in \Zb} (-1)^{l \sum_i m_i} q^{\frac{l}{2}\sum_i m_i^2} z^{-l \sum_i m_i}
  \nonumber \\
&\times \left( \prod_{i \ne j}^N \frac{(q^{\frac{3}{4} + m_i - m_j} t^{-1} x^{\pm} u_i u_j^{-1}; q)_{\infty}}{(q^{1 + m_i - m_j} u_i u_j^{-1}; q)_{\infty} (q^{\frac{1}{2} + m_i - m_j} t^{-2} u_i u_j^{-1}; q)_{\infty}} \right)
\prod_{i = 1}^N u_i^{l m_i} \prod_{\alpha = 1}^l
(q^{\frac34 \pm m_i}t^{-1}u_i^{\pm} y_{\alpha}^{\pm};q)_{\infty} \; . 
\end{align}
Here $u_i$ are the fugacities for the boundary global symmetry associated with the broken gauge group 
and $z$ the fugacity for the topological symmetry. 

In the Abelian case, $N=1$, the half-index is 
\begin{align}
\label{h_ADHMu1nfl_DDD}
\mathbb{II}_{\mathcal{D},D,D}^{\textrm{$U(1)$ ADHM}-[l]}
& = \frac{(q^{\frac{3}{4}} t^{-1} x^{\pm}; q)_{\infty}}{(q)_{\infty} (q^{\frac12}t^{-2};q)_{\infty}} \sum_{m \in \Zb} (-1)^{l m} q^{\frac{l}{2} m^2} z^{-l m} u_1^{lm}
\prod_{\alpha = 1}^l (q^{\frac34 \pm m}t^{-1}u_1^{\pm} y_{\alpha}^{\pm};q)_{\infty} \; .
\end{align}

Let us consider the special fugacity limits of the half-index (\ref{h_ADHMuNnfl_DDD}). 
When we take the C-twist limit $t\rightarrow q^{-1/4}$ \cite{Gaiotto:2019jvo,Okazaki:2019bok} with $x=z=1$, 
the half-index (\ref{h_ADHMuNnfl_DDD}) reduces to 
\begin{align}
\label{h_ADHMuNnfl_DDD_C}
&
\mathbb{II}_{\mathcal{D},D,D}^{{\textrm{$U(N)$ ADHM}-[l]}^{(C)}}(u_i,y_{\alpha};q)
\nonumber\\
&=\frac{1}{(q)_{\infty}^N}
\sum_{m_i\in \mathbb{Z}}
(-1)^{l\sum_{i=1}^N m_i}
q^{\frac{l}{2}\sum_{i=1}^N m_i^2}
\prod_{i\neq j}^N
\frac{1}{(q^{1+m_i-m_j}u_iu_j^{-1};q)_{\infty}}
\prod_{i=1}^{N} u_i^{lm_i}
\prod_{\alpha = 1}^l  (q^{1\pm m_i}u_i^{\pm} y_{\alpha}^{\pm};q)_{\infty} \; . 
\end{align}

We will now further simplify by setting all fugacities $u_i$ to unity. We do not have a general proof but conjecture, supported by numerical calculations, that the half-index remains well-defined as a $q$-series in this limit, i.e.\ the coefficient of each power of $q$ remains finite\footnote{This is obvious for $N = 1$ and can also be checked analytically for $N = 2$.}. With this assumption it is useful to define
\begin{align}
f_m(X) & = \left \{ \begin{array}{lll} 1 & , & m = 0 \\ 1 - X^{sgn(m)} & , & m \ne 0 \end{array} \right.
\end{align}
to rewrite \eqref{h_ADHMuNnfl_DDD_C} as, after setting $y_{\alpha} = 1$,
\begin{align}
\label{h_ADHMuNnfl_DDD_C_zeros}
\mathbb{II}_{\mathcal{D},D,D}^{{\textrm{$U(N)$ ADHM}-[l]}^{(C)}}(u_i; q)
 = &
\frac{1}{(q)_{\infty}^N}
\sum_{m_i\in \mathbb{Z}}
(-1)^{(N-1)\sum_{i=1}^N m_i}
q^{\frac{l}{2}\sum_{i=1}^N |m_i| + \frac{1}{4}\sum_{i, j = 1}^N \left( (m_i - m_j)^2 - |m_i - m_j| \right)}
 \nonumber \\
 &\times \frac{\prod_{i=1}^{N} u_i^{Nm_i - \sum_{j = 1}^N m_j}
(q u_i^{\pm};q)_{\infty}^l f_{m_i}(u_i)^l}{\prod_{i > j}^N (q u_i u_j^{-1};q)_{\infty} f_{m_i - m_j}(u_i u_j^{-1})} \; . 
\end{align}

When we set the global fugacities $u_i$ to unity, we see that we get zeros in both the numerator and the denominator. With some simple rearrangement, if exactly $k$ of the $m_i$ are non-zero, these take the form of $kl$ factors of $(1 - u_i)$ in the numerator and between $k(N - k)$ and $k(N - k) + \frac{1}{2}k(k-1)$ factors of $(u_i - u_j)$ in the denominator where this number depends on how many of the $k$ non-zero $m_i$ are distinct. When taking the limit $u_i \to 1$ we see that we get an overall factor of $(q)_{\infty}^{N(2l - N)}$. In the sum over $m_i$ we see that, other than this overall factor of $(q)_{\infty}^{N(2l - N)}$ the $q$ dependence is given by a power of $q$ which is invariant under permutations of the $m_i$ as well as reversing the sign of all $m_i$. The terms in the half-index which are not invariant under such transformations are given by the factor
\begin{align}
    X_{\vec{m}} = & \frac{\prod_{i=1}^{N} u_i^{Nm_i - \sum_{j = 1}^N m_j}
 f_{m_i}(u_i)^l}{\prod_{i > j}^N f_{m_i - m_j}(u_i u_j^{-1})} \; .
\end{align}
Letting $[\vec{m}]$ denote the equivalence class of $\vec{m}$ under permutation of components and simultaneous reversal of sign of all components, and $Y_{[\vec{m}]}$ the sum of $X_{\vec{m}}$ over elements of such equivalence classes with integer components we can write
\begin{align}
\mathbb{II}_{\mathcal{D},D,D}^{{\textrm{$U(N)$ ADHM}-[l]}^{(C)}}(u_i;q)
 = &
\frac{1}{(q)_{\infty}^N}
\sum_{[\vec{m}]}
(-1)^{(N-1)\sum_{i=1}^N m_i}
q^{\frac{l}{2}\sum_{i=1}^N |m_i| + \frac{1}{4}\sum_{i, j = 1}^N \left( (m_i - m_j)^2 - |m_i - m_j| \right)}
 \nonumber \\
 &\times \frac{\prod_{i=1}^{N} (q u_i^{\pm};q)_{\infty}^l}{\prod_{i > j}^N (q u_i u_j^{-1};q)_{\infty}} Y_{[\vec{m}]} \; . 
\end{align}
Assuming that whenever the half-index is well defined in the limit $u_i \to 1$ that $Y_{[\vec{m}]}$ is for all $\vec{m}$, with limit $\widehat{Y}_{[\vec{m}]}$, 
we see that this limit will give 
\begin{align}
&
\widehat{\mathbb{II}}_{\mathcal{D},D,D}^{{\textrm{$U(N)$ ADHM}-[l]}^{(C)}}(q)
\nonumber\\
&=
(q)_{\infty}^{N(2l-N)}
\sum_{[\vec{m}]}
(-1)^{(N-1)\sum_{i=1}^N m_i}
q^{\frac{l}{2}\sum_{i=1}^N |m_i| + \frac{1}{4}\sum_{i, j = 1}^N \left( (m_i - m_j)^2 - |m_i - m_j| \right)} \widehat{Y}_{[\vec{m}]} \; . 
\end{align}

In the sum over $X_{\vec{m}}$ additional cancellations can occur but if $\widehat{Y}_{[\vec{m}]}$ is well defined and $kl > k(N - k) + \frac{1}{2}k(k-1) = k(N - \frac{k+1}{2})$ then clearly we must have $\widehat{Y}_{[\vec{m}]} = 0$. 
It follows that for $l\ge N$, the only contribution to $\widehat{\mathbb{II}}$ is the term with all $m_i=0$ and hence 
\begin{align}
\widehat{\mathbb{II}}_{\mathcal{D},D,D}^{{\textrm{$U(N)$ ADHM}-[l \ge N]}^{(C)}}(q) = & (q)_{\infty}^{2Nl-N^2}
\end{align}

For $l<N$ the result involves contributions with some or all $m_i\neq 0$ for $kl \le k(N - \frac{k+1}{2})$. 
For example, when $l = N-1$ we have non-zero contributions only for $k = 0$ or $k = 1$. Specifically, for $N=2$ and $l=1$, we find that indeed $\widehat{Y}_{[\vec{m}]} = 0$ for $k = 2$ and that $\widehat{Y}_{[\vec{m}]} = 2$ for $k = 1$ while in all cases obviously $\widehat{Y}_{[\vec{m}]} = 1$ for $k = 0$ so we find
\begin{align}
\widehat{\mathbb{II}}_{\mathcal{D},D,D}^{{\textrm{$U(2)$ ADHM}-[1]}^{(C)}}(q)
&=\sum_{m\in \mathbb{Z}}(-1)^m q^{\frac{m^2}{2}}=\vartheta_4(q). 
\end{align}

In this case, the half-index is non-trivial in the Coulomb limit \cite{Razamat:2014pta} where we send $q \to 0$ with $\mathfrak{t} = q^{\frac{1}{4}} t^{-1}$ fixed. 
This gives
\begin{align}
\label{h_ADHMuNnfl_DDD_Coulomb}
&
\mathbb{II}_{\mathcal{D},D,D \; \textrm{Coulomb}}^{\textrm{$U(N)$ ADHM}-[l]}
\nonumber\\
&= \lim_{q \to 0} \frac{1}{(\mathfrak{t}^2; q)_{\infty}^N} \sum_{m_i \in \Zb} (-1)^{l \sum_i m_i} q^{\frac{l}{2}\sum_i m_i^2} z^{-l \sum_i m_i}
  \nonumber \\
&\times \left( \prod_{i \ne j}^N \frac{(q^{\frac{1}{2} + m_i - m_j} \mathfrak{t} x^{\pm} u_i u_j^{-1}; q)_{\infty}}{(q^{1 + m_i - m_j} u_i u_j^{-1}; q)_{\infty} (q^{m_i - m_j} \mathfrak{t}^2 u_i u_j^{-1}; q)_{\infty}} \right)
\prod_{i = 1}^N
(q^{\frac12 \pm m_i} \mathfrak{t} u_i^{\pm};q)_{\infty}^l
u_i^{l m_i}
\nonumber \\
& = \left( \prod_{i, j = 1}^N \frac{1}{(1 - \mathfrak{t}^2 u_i u_j^{-1})} \right) \sum_{m_i \in \Zb} z^{-l \sum_i m_i} \mathfrak{t}^{l \sum_i |m_i|} \prod_{i > j} \frac{F_{m_i - m_j}(u_i u_j^{-1}, \mathfrak{t}^2)}{F_{m_i - m_j}(u_i u_j^{-1}, 1)}, 
\end{align}
where
\begin{align}
F_m(X, Y) & = \left \{ \begin{array}{lll} 1 & , & m = 0 \\ 1 - X^{sgn(m)}Y & , & m \ne 0 \end{array} \right. \; .
\end{align}

For $N = 1$ we have a particularly simple result
\begin{align}
\label{h_ADHMu1nfl_DDD_Coulomb}
\mathbb{II}_{\mathcal{D},D,D \; \textrm{Coulomb}}^{\textrm{$U(1)$ ADHM}-[l]}
& = \frac{1}{(1 - \mathfrak{t}^2)} \sum_{m \in \Zb} z^{-lm} \mathfrak{t}^{l|m|}
 = \frac{1}{(1 - \mathfrak{t}^2)} \left( \frac{1}{(1 - z^l \mathfrak{t}^l)} + \frac{1}{(1 - z^{-l} \mathfrak{t}^l)} - 1 \right)
 \nonumber \\
 & = \frac{1 - \mathfrak{t}^{2l}}{(1 - \mathfrak{t}^2)(1 - z^l \mathfrak{t}^l) (1 - z^{-l} \mathfrak{t}^l)} \; .
\end{align}
The expression (\ref{h_ADHMu1nfl_DDD_Coulomb}) agrees with the Coulomb index of the $U(1)$ ADHM theory, 
that is the refined Hilbert series for the $\mathbb{C}^2/\mathbb{Z}_l$ \cite{Cremonesi:2013lqa}. 
For $N>1$ the Coulomb limit (\ref{h_ADHMuNnfl_DDD_Coulomb}) is different from the Coulomb index of the $U(N)$ ADHM theory.

For $N = 2$ we find
\begin{align}
\label{h_ADHMu2nfl_DDD_Coulomb}
&
\mathbb{II}_{\mathcal{D},D,D \; \textrm{Coulomb}}^{\textrm{$U(2)$ ADHM}-[l]}
\nonumber\\
& = \left( \prod_{i, j = 1}^2 \frac{1}{(1 - \mathfrak{t}^2 u_i u_j^{-1})} \right) \sum_{m_i \in \Zb} z^{-l \sum_i m_i} \mathfrak{t}^{l \sum_i |m_i|} \frac{F_{m_2 - m_1}(u_2 u_1^{-1}, \mathfrak{t}^2)}{F_{m_2 - m_1}(u_2 u_1^{-1}, 1)} \; ,
\end{align}
where
\begin{align}
\frac{F_{m_2 - m_j}(u_2 u_1^{-1}, \mathfrak{t}^2)}{F_{m_2 - m_1}(u_2 u_1^{-1}, 1)} & = \left \{ \begin{array}{lll} \frac{1 - \mathfrak{t}^2 u_2 u_1^{-1}}{1 - u_2 u_1^{-1}} & , & m_2 > m_1 \\1 & , & m_2 = m_1 \\ \frac{1 - \mathfrak{t}^2 u_1 u_2^{-1}}{1 - u_1 u_2^{-1}} & , & m_2 < m_1 \end{array} \right. \; .
\end{align}

\subsection{ADHM quiver theory}
Next we present half-indices for the basic $\mathcal{N}=(0,4)$ half-BPS boundary conditions in the ADHM quiver theories 
with either Neumann or Dirichlet boundary conditions for the vector multiplets.
For simplicity we have not included the contributions from the topological fugacities $y_I$ but these can easily be restored if desired.

\subsubsection{ADHM quiver theory with $(\mathcal{N},N,N)+\Gamma+\eta$}
Let us consider the case of Neumann boundary conditions $(\mathcal{N},N,N)+\Gamma+\eta$. 
The half-index is given by the matrix integral 
\begin{align}
\label{ADHMquiveruN_04Neu}
&
\mathbb{II}_{\mathcal{N},N,N+\Gamma+\eta}^{\textrm{$U(N)^l$ ADHM quiver}}(t,x,z,u_1;q)
\nonumber\\
 = & \frac{1}{(N!)^l}(q)_{\infty}^{Nl} (q^{\frac12}t^{-2};q)_{\infty}^{Nl} 
\prod_{I = 1}^l \oint \prod_{i=1}^{N}\frac{ds^{(I)}_i}{2\pi is^{(I)}_i}
\prod_{i<j}
(s_i^{(I) \pm}s_j^{(I) \mp};q)_{\infty}
(q^{\frac12}t^{-2}s_i^{(I) \pm}s_j^{(I) \mp};q)_{\infty}
\nonumber\\
&\times \prod_{i,j}
\frac{1}
{(q^{\frac14}t^{-1}s_i^{(I) \pm}s_j^{(I+1) \mp}z^{\mp};q)_{\infty}}
\left(q^{\frac12}(\prod_i s_i^{(I)})^{\pm} (\prod_j s_j^{(I+1)})^{\mp} u_1^{\pm} z^{\mp};q\right)_{\infty}
\nonumber \\
 &\times \prod_{i=1}^N \frac{1}
{(q^{\frac14}t^{-1}s_i^{(1) \pm};q)_{\infty}}
\prod_{i=1}^N (q^{\frac12}s_i^{(1) \pm}x^{\pm};q)_{\infty} \; ,
\end{align}
where we note that since we have a circular quiver we identify indices $I = l+1$ with $I = 1$.

In the Abelian case $N = 1$, the half-index reads
\begin{align}
\label{ADHMquiveru1_04Neu}
\mathbb{II}_{\mathcal{N},N,N+\Gamma+\eta}^{\textrm{$U(1)^l$ ADHM quiver}}(t,x,z,u_1;q)
&= (q)_{\infty}^{l} (q^{\frac12}t^{-2};q)_{\infty}^{l} 
\prod_{I = 1}^l \oint \frac{ds^{(I)}}{2\pi is^{(I)}}
\frac{1}
{(q^{\frac14}t^{-1}s^{(I) \pm}s^{(I+1) \mp}z^{\mp};q)_{\infty}}
\nonumber\\
 &\times (q^{\frac12} s^{(I) \pm} s^{(I+1) \mp} u_1^{\pm} z^{\mp};q)_{\infty}
\frac{1}
{(q^{\frac14}t^{-1}s^{(1) \pm};q)_{\infty}}
(q^{\frac12}s^{(1) \pm}x^{\pm};q)_{\infty}. 
\end{align}

In particular, for $l=1$ the theory can be viewed as the mirror description of the $U(1)$ ADHM theory 
which can be constructed by swapping the supermultiplets with the twisted supermultiplets. 
Thus the expression simply results from (\ref{ADHMu1_1_04Neu}) upon the action of mirror symmetry, 
$t$ $\rightarrow$ $t^{-1}$, $x\rightarrow z$ and $z\rightarrow x$. 
For $l=2$ we have
\begin{align}
\label{ADHMquiveru1_2_04Neu}
&
\mathbb{II}_{\mathcal{N},N,N+\Gamma+\eta}^{\textrm{$U(1)^2$ ADHM quiver}}(t,x,z,u_1;q)
\nonumber\\
&= (q)_{\infty}^{2} (q^{\frac12}t^{-2};q)_{\infty}^{2} 
\oint \frac{ds^{(1)}}{2\pi is^{(1)}}
\oint \frac{ds^{(2)}}{2\pi is^{(2)}}
\nonumber \\
&\times \frac{(q^{\frac12} s^{(1) \pm} s^{(2) \mp} u_1^{\pm} z^{\mp};q)_{\infty}
 (q^{\frac12} s^{(2) \pm} s^{(1) \mp} u_1^{\pm} z^{\mp};q)_{\infty}}
{(q^{\frac14}t^{-1}s^{(1) \pm}s^{(2) \mp}z^{\mp};q)_{\infty} (q^{\frac14}t^{-1}s^{(2) \pm}s^{(1) \mp}z^{\mp};q)_{\infty}}
\frac{(q^{\frac12}s^{(1) \pm}x^{\pm};q)_{\infty}}
{(q^{\frac14}t^{-1}s^{(1) \pm};q)_{\infty}}. 
\end{align}

Let us consider the special fugacity limits of the Neumann half-index (\ref{ADHMquiveruN_04Neu}). 
As we will see in section \ref{sec_Dualities}, the Neumann boundary condition in the Abelian ADHM quiver theory is 
dual to the Dirichlet boundary condition in the Abelian ADHM theory. 
So the special fugacity limits for the half-index (\ref{ADHMquiveru1_04Neu}) with $N=1$ 
are exactly same as those for the Dirichlet half-index in the Abelian ADHM theory, which we already discussed. 
On the other hand, for the non-Abelian case $N>1$, the Neumann boundary condition in the ADHM quiver theory is not dual to the Dirichlet boundary condition. 

Taking the C-twist limit \cite{Gaiotto:2019jvo,Okazaki:2019bok} with $x=z=u_1=1$, 
the Neumann half-index (\ref{ADHMquiveruN_04Neu}) becomes 
\begin{align}
&
\mathbb{II}_{\mathcal{N},N,N+\Gamma+\eta}^{\textrm{$U(N)^l$ ADHM quiver}^{(C)}}(q)
\nonumber\\
&=\frac{1}{(N!)^l}
(q)_{\infty}^{2Nl}\prod_{I=1}^l 
\oint \prod_{i=1}^N 
\frac{ds_i^{(I)}}{2\pi is_i^{(I)}}
\prod_{i<j}(s_i^{(I)\pm}s_j^{(I)\mp};q)_{\infty}(qs_i^{(I)\pm}s_j^{(I)\mp};q)_{\infty}
\nonumber\\
&\times 
\prod_{i,j}
\frac{1}{(q^{\frac14}s_{i}^{(I)\pm}s_j^{(I+1)\mp};q)_{\infty}}
\left(q^{\frac12}(\prod_i s_i^{(I)})^{\pm} (\prod_j s_j^{(I+1)})^{\mp};q\right)_{\infty}. 
\end{align}
Note that if we had not included the contribution from the bi-determinant Fermi $\eta$, 
this is identical to the unflavored Schur index of $\mathcal{N}=2$ circular quiver $U(N)^l$ gauge theory. 
Though we do not here pursue further, it would be interesting to find the exact closed-form expression 
by employing the Fermi-gas method, as argued in \cite{Bourdier:2015sga} for the unflavored Schur index. 

When we take the Coulomb limit \cite{Razamat:2014pta} where we set $q \to 0$ with $\mathfrak{t} = q^{\frac{1}{4}} t^{-1}$ fixed, 
we find that the matrix integral (\ref{ADHMquiveruN_04Neu}) reduces to the Higgs index for the ADHM quiver theory. 
According to mirror symmetry it is equivalent to the Coulomb index of the $U(N)$ ADHM theory, 
that is the refined Hilbert series for $\mathrm{Sym}^N(\mathbb{C}^2/\mathbb{Z}_l)$ \cite{Cremonesi:2013lqa}. 

\subsubsection{ADHM $U(N)^l$ quiver theory with $(\mathcal{D},D,D)$}
Next consider the case of Dirichlet boundary conditions $(\mathcal{D},D,D)$ in the ADHM quiver theory. 
The half-index is evaluated as the following infinite series: 
\begin{align}
\label{h_ADHMquiver_uN_l_DDD}
&
\mathbb{II}_{\mathcal{D},D,D}^{\textrm{$U(N)^l$ ADHM quiver}}
(t,x,z,u_I;q)
\nonumber\\
&= \frac{1}{(q)_{\infty}^{lN} (q^{\frac12}t^2;q)_{\infty}^{lN}} \sum_{m^{(I)}_i \in \Zb} (-1)^{\sum_{i=1}^N m^{(I)}_i}
\nonumber \\
&\times q^{\frac{3}{2} \sum_{i=1}^N (m^{(1)}_i)^2 + \sum_{I = 2}^l \sum_{i=1}^N (m^{(I)}_i)^2 - \sum_{I = 1}^l \sum_{i, j=1}^N m^{(I)}_i m^{(I+1)}_j} x^{-\sum_{i=1}^N m^{(1)}_i}
\nonumber \\
 &\times \left( \prod_{I = 1}^l \prod_{i=1}^N u_{Ii}^{\sum_{j=1}^N (2m^{(I)}_j - m^{(I-1)}_j - m^{(I)+1}_j)}
(q^{\frac34 \pm m_I \mp m_{I+1}}t u_I^{\pm} u_{I+1}^{\mp} z^{\mp};q)_{\infty}
\right)
\left( \prod_{i=1}^N u_{1i}^{m^{(1)}_i} \right)
\nonumber \\
&\times \left( \prod_{I = 1}^l \prod_{i \ne j}^N 
\frac{1}{(q^{1 + m^{(I)}_i - m^{(I)}_j} u_i u_j^{-1}; q)_{\infty} (q^{\frac{1}{2} + m^{(I)}_i - m^{(I)}_j} t^{2} u_i u_j^{-1}; q)_{\infty}} \right)
\nonumber \\
&\times \left( \prod_{i, j=1}^N (q^{\frac{3}{4} + m^{(I)}_i - m^{(I+1)}_j} t z^{\pm} u_{Ii} u_{(I+1)j}^{-1}; q)_{\infty} \right)
\prod_{i=1}^N (q^{\frac34 \pm m^{(1)}_i}t u_{1i}^{\pm};q)_{\infty}. 
\end{align}

For example, in the Abelian case we have the half-index of the form 
\begin{align}
\label{h_ADHMquiver_u1_l_DDD}
&
\mathbb{II}_{\mathcal{D},D,D}^{\textrm{$U(1)^l$ ADHM quiver}}
\nonumber\\
& = \frac{1}{(q)_{\infty}^l (q^{\frac12}t^2;q)_{\infty}^l} \sum_{m_I \in \Zb} (-1)^{m_1} q^{\frac{3}{2} m_1^2 + \sum_{I = 2}^l m_I^2 - \sum_{I = 1}^l m_I m_{I+1}} x^{-m_1}
\nonumber \\
&\times \left( \prod_{I = 1}^l u_I^{2m_I - m_{I-1} - m_{I+1}}
(q^{\frac34 \pm m_I \mp m_{I+1}}t u_I^{\pm} u_{I+1}^{\mp} z^{\mp};q)_{\infty}
\right)
u_1^{m_1}
(q^{\frac34 \pm m_1}t u_1^{\pm};q)_{\infty} \; .
\end{align}

Again for $l=1$ the ADHM and ADHM quiver theories are the same 
so the Dirichlet half-indices are identical to (\ref{h_ADHMu1nfl_DDD}) with $l=1$ 
up to relabelling fugacities under the mirror map $t \to -t$ and $x \leftrightarrow z$. 
For $l=2$ we have
\begin{align}
\label{h_ADHMquiver_u1_2_DDD}
&
\mathbb{II}_{\mathcal{D},D,D}^{\textrm{$U(1)^2$ ADHM quiver}}
\nonumber\\
& = \frac{1}{(q)_{\infty}^2 (q^{\frac12}t^2;q)_{\infty}^2} \sum_{m_I \in \Zb} (-1)^{m_1} q^{\frac{3}{2} m_1^2 + m_2^2 - 2 m_1 m_2} x^{-m_1}
u_1^{2m_1 - 2m_2} u_2^{2m_2 - 2m_1}
\nonumber \\
 & \times 
(q^{\frac34 \pm m_1 \mp m_2} t u_1^{\pm} u_2^{\mp} z^{\mp};q)_{\infty}
(q^{\frac34 \pm m_1 \mp m_2} t u_1^{\pm} u_2^{\mp} z^{\pm};q)_{\infty}
(q^{\frac34 \pm m_1}t u_1^{\pm};q)_{\infty} \; .
\end{align}
As discussed in section \ref{sec_Dualities}, 
the Dirichlet half-index (\ref{h_ADHMquiver_u1_l_DDD}) for $N=1$ coincides with the Neumann half-index in the $U(N)$ ADHM theory 
as a consequence of dualities of boundary conditions. 
In the H-twist limit \cite{Gaiotto:2019jvo,Okazaki:2019bok} with $x=z=1$, 
the Dirichlet half-index (\ref{h_ADHMquiver_uN_l_DDD}) reduces to
\begin{align}
\label{h_ADHMquiver_uN_l_DDD_Htwist}
&
\mathbb{II}_{\mathcal{D},D,D}^{\textrm{$U(N)^l$ ADHM quiver}^{(H)}}
\nonumber\\
&= \frac{1}{(q)_{\infty}^{2lN}} \sum_{m^{(I)}_i \in \Zb} (-1)^{\sum_{i=1}^N m^{(I)}_i} 
\prod_{I = 1}^l \prod_{i=1}^N u_{Ii}^{\sum_{j=1}^N (2m^{(I)}_j - m^{(I-1)}_j - m^{(I)+1}_j)}
u_{1i}^{m^{(1)}_i}
\nonumber \\
&\times q^{\frac{3}{2} \sum_{i=1}^N (m^{(1)}_i)^2 + \sum_{I = 2}^l \sum_{i=1}^N (m^{(I)}_i)^2 - \sum_{I = 1}^l \sum_{i, j=1}^N m^{(I)}_i m^{(I+1)}_j} 
\nonumber \\
&\times \prod_{I = 1}^l \prod_{i \ne j}^N 
\frac{1}{(q^{1\pm m^{(I)}_i \mp m^{(I)}_j} u_{Ii} ^{\pm} u_{Ij}^{\mp}; q)_{\infty}} 
\nonumber \\
&\times \prod_{I = 1}^l \prod_{i, j=1}^N (q^{1\pm m^{(I)}_i \mp m^{(I+1)}_j} u_{Ii}^{\pm} u_{(I+1)j}^{\mp}; q)_{\infty} 
\prod_{i=1}^N (q^{1\pm m^{(1)}_i}u_{1i}^{\pm};q)_{\infty}. 
\end{align} 

The Dirichlet half-index (\ref{h_ADHMquiver_uN_l_DDD}) for the ADHM quiver theory also admits 
a non-trivial Higgs limit \cite{Razamat:2014pta}. 

\subsection{$U(1)_k \times U(1)_{-k}$ ABJM theory}
In this section we consider the half-indices for a range of possible Abelian ABJM theory boundary conditions.
We work in an $\Ncal = 2$ description so allow $\mathcal{N}=(0,2)$ Neumann or Dirichlet boundary conditions \cite{Okazaki:2013kaa}  
independently for the four $\Ncal = 2$ chiral multiplets $H$, $\tilde{H}$, $T$ and $\tilde{T}$ 
coming from the $\Ncal = 4$ hypermultiplet and twisted hypermultiplet. 
For the vector multiplets we consider the boundary conditions on the linear combinations of the original $U(1)$ gauge groups discussed in section~\ref{Sec_ABJM_U1k_diag}. 
Half-indices for the boundary conditions imposed on the original $U(1)$ gauge groups are presented in Appendix~\ref{ABJM_U1k_VMbc}.

\subsubsection{Diagonal gauge group breaking -- $\Zb_k$ gauge theory}
We consider the $\mathcal{N}=(0,2)$ boundary conditions for the vector multiplets \cite{Okazaki:2013kaa} as in section~\ref{Sec_ABJM_U1k_diag}.
For the $U(1)_{-}$ vector multiplet with $\mathcal{N}=(0,2)$ Dirichlet boundary condition 
we have the contribution $1/(q)_{\infty}$ but since the mixed gauge anomaly breaks the $U(1)_{u_{-}}$ global symmetry 
we set the fugactity $u_{-} = 1$ and do not have a sum over monopole flux. 
We have the usual contribution from the $U(1)_{+}$ vector multiplet with Neumann boundary conditions 
but note that nothing is charged under this gauge group. Altogether this appears to give the contribution to the half-index
\begin{align}
    \II^{U(1)_{+} \times U(1)_{-} \textrm{ ABJM}_k}_{\Ncal, \Dcal + \Gamma} = & \frac{1}{(q)_{\infty}} (q)_{\infty} \oint \frac{ds_{+}}{2 \pi i s_{+}}
    \nonumber \\
    = & 1, 
\end{align}
which multiplies the contributions from any 2d chirals or Fermis and the hypermultiplet and twisted hypermultiplet, noting that these are singlets under the $U(1)_{+}$ gauge group.
I.e.\ this would just give the half-index for the 4 free chirals $H$, $\tilde{H}$, $T$ and $\tilde{T}$ along with any 2d matter contributions.

However, as argued in section~\ref{Sec_ABJM_U1k_diag} 
we claim that the result is a system with residual $\Zb_k$ gauge group on the boundary which couples to the 3d chiral multiplets.

We also assume that we have the bi-determinant Fermi, which in the Abelian case considered here has $U(1)_k \times U(1)_{-k}$ charges $(+1, -1)$ as well as charge $+1$ under the global $U(1)_u$ and $-1$ under the global $U(1)_x$. The charge under the gauge group is then charge $+1$ under $U(1)_{-}$ and uncharged under $U(1)_{+}$. This reduces to being charged under the residual $\Zb_k$ gauge group only.

Altogether this gives the contribution to the half-index
\begin{align}
\label{ABJM_hind}
    \II^{U(1)_k \times U(1)_{-k} \textrm{ ABJM}}_{\Ncal, \Dcal + \Gamma} = & 
    \sum_{\lambda = 0}^{k-1}
    (q^{\frac12} \omega^{\pm \lambda} u^{\pm}x^{\mp};q)_{\infty} \frac{1}{(q)_{\infty}} (q)_{\infty} \oint \frac{ds_{+}}{2 \pi i s_{+}}
    \II^{\textrm{3d chirals}}
    \nonumber \\
    = & \sum_{\lambda = 0}^{k-1} (q^{\frac12} \omega^{\pm \lambda} u^{\pm}x^{\mp};q)_{\infty}
    \II^{\textrm{3d chirals}}, 
\end{align}
where $\omega = \exp(2 \pi i/k)$ and $\II^{\textrm{3d chirals}}$ denotes the contributions 
from the hypermultiplet and twisted hypermultiplet, noting that these are singlets under the $U(1)_{+}$ gauge group. 
These contributions are summarized in the following 
where we can freely choose Neumann or Dirichlet boundary conditions for each of the chiral multiplets $H$, $\tilde{H}$, $T$, $\tilde{T}$.
\begin{align}
    \begin{array}{c|c|c}
     & N & D \\ \hline
    H & (q^{\frac14} \omega^{\lambda} x t;q)_{\infty}^{-1} & (q^{\frac34} \omega^{-\lambda} x^{-1} t^{-1};q)_{\infty} \\
    \tilde{H} & (q^{\frac14} \omega^{-\lambda} x^{-1} t;q)_{\infty}^{-1} & (q^{\frac34} \omega^{\lambda} x t^{-1};q)_{\infty} \\
    T & (q^{\frac14} \omega^{\lambda} z t^{-1};q)_{\infty}^{-1} & (q^{\frac34} \omega^{-\lambda} z^{-1} t;q)_{\infty} \\
    \tilde{T} & (q^{\frac14} \omega^{-\lambda} z^{-1} t^{-1};q)_{\infty}^{-1} & (q^{\frac34} \omega^{\lambda} z t;q)_{\infty}
    \end{array}
\end{align}
Note that in the case $k=1$ we have no discrete gauge group so there is no sum, just the contributions from free 3d chirals.

\subsection{$U(1)_1 \times U(1)_0^{l-1} \times U(1)_{-1}$ circular quiver CS theory}
We now present half-indices for the Abelian circular quiver Chern-Simons theories. In all cases we take the diagonal breaking of the $U(1)_1 \times U(1)_{-1}$ gauge group. For the remaining $U(1)^{l-1}$ twisted vector multiplets, the hypermultiplet and the $l$ twisted hypermultiplets we take boundary conditions $(\Ncal, D, N)$ along with Fermis $\eta_{I, I+1}$ or $(\Dcal, N, D)$ along with Fermi $\eta$.
As for the ADHM quiver theories, for simplicity we have set the topological fugacities $y_I = 1$.

In the case of $(\Dcal, N, D) + \eta$ boundary conditions we have half-index
\begin{align}
\label{QuiverCS_04Dir}
&
\mathbb{II}_{(\Dcal,N,D)+\eta}^{\textrm{$U(1)^{l+1}$ Quiver CS}}(t,x,z,u_I;q)
\nonumber\\
 &=  \frac{1}{(q)_{\infty}^{l-1} (q^{\frac12}t^2;q)_{\infty}^{l-1}}
\oint \frac{ds_{+}}{2\pi i s_{+}} 
\sum_{m_I \in \Zb} q^{\frac{1}{2}\sum_{I = 1}^l m_I^2} (s_{+}zux^{-1})^{\sum_{I = 1}^l m_I}
  \nonumber \\
 &\times \frac{(q^{\frac{1}{2} \pm \sum_{I = 1}^l m_I} \left( \prod_{I = 1}^l v_I \right)^{\pm} u^{\pm} x^{\mp}; q)_{\infty}}{(q^{\frac14  \pm \sum_{I = 1}^l m_I} \left( \prod_{I = 1}^l v_I \right)^{\pm} tx^{\pm};q)_{\infty}}
\prod_{I = 1}^l v_I^{m_I} 
 (q^{\frac{3}{4} \pm m_I} t v_I^{\pm} z^{\pm}; q)_{\infty} \; .
\end{align}
It is trivial to evaluate the contour integral leading to the expression
\begin{align}
\label{QuiverCS_04Dir_int}
&
\mathbb{II}_{(\Dcal,N,D)+\eta}^{\textrm{$U(1)^{l+1}$ Quiver CS}}(t,x,z,u_I;q)
\nonumber\\
 &=  \frac{1}{(q)_{\infty}^{l-1} (q^{\frac12}t^2;q)_{\infty}^{l-1}}
\sum_{m_I \in \Zb \vert \sum_{I = 1}^l m_I = 0} q^{\frac{1}{2}\sum_{I = 1}^l m_I^2} 
  \nonumber \\
 &\times \frac{(q^{\frac{1}{2}} \left( \prod_{I = 1}^l v_I \right)^{\pm} u^{\pm} x^{\mp}; q)_{\infty}}{(q^{\frac14} \left( \prod_{I = 1}^l v_I \right)^{\pm} tx^{\pm};q)_{\infty}}
\prod_{I = 1}^l v_I^{m_I} 
 (q^{\frac{3}{4} \pm m_I} t v_I^{\pm} z^{\pm}; q)_{\infty}
\end{align}
and taking account of the total breaking of the $U(1)_{-}$ gauge group which has fugacity $\prod_{I = 1}^l v_I$ we should specialize the fugacities by setting $\prod_{I = 1}^l v_I = 1$, which of course is consistent with the constraint $\sum_{I = 1}^l m_I = 0$, to arrive at
\begin{align}
\label{QuiverCS_04Dir_int_red}
&
\mathbb{II}_{(\Dcal,N,D)+\eta}^{\textrm{$U(1)^{l+1}$ Quiver CS}}(t,x,z,u_I;q)
\nonumber\\
 &=  \frac{1}{(q)_{\infty}^{l-1} (q^{\frac12}t^2;q)_{\infty}^{l-1}}
\sum_{m_I \in \Zb \vert \sum_{I = 1}^l m_I = 0} q^{\frac{1}{2}\sum_{I = 1}^l m_I^2} 
  \nonumber \\
 &\times \frac{(q^{\frac{1}{2}} u^{\pm} x^{\mp}; q)_{\infty}}{(q^{\frac14} tx^{\pm};q)_{\infty}}
\prod_{I = 1}^l v_I^{m_I} 
 (q^{\frac{3}{4} \pm m_I} t v_I^{\pm} z^{\pm}; q)_{\infty}, 
\end{align}
which is also what we would have found starting from the reduced field content \eqref{U1_l_CSQuiver_charges_reduced}.

In the case of $(\Ncal, D, N) + \eta_{I}$ boundary conditions we have half-index
\begin{align}
\label{QuiverCS_04Neu_int_red}
&
\mathbb{II}_{(\Ncal,D,N)+\eta_{I, I+1}}^{\textrm{$U(1)^{l+1}$ Quiver CS}}(t,x,z,u_I;q)
\nonumber\\
&=  (q)_{\infty}^{l-1} (q^{\frac12}t^{-2};q)_{\infty}^{l-1}
(q^{\frac34} t^{-1} x^{\pm};q)_{\infty}
  \nonumber \\
 &\times \left( \prod_{I = 1}^{l-1} \frac{ds_I}{2\pi i s_I} \right)
 \prod_{I = 1}^l \frac{(q^{\frac{1}{2}} s_I^{\pm} u^{\pm} z^{\mp}; q)_{\infty}}{(q^{\frac{1}{4}} s_I^{\pm} z^{\pm} t^{-1}; q)_{\infty}}, 
\end{align}
where $s_l = \prod_{I = 1}^{l-1} s_I^{-1}$.

\section{Dualities of boundary conditions}
\label{sec_Dualities}




We now present various examples of dualities between Abelian ADHM, ADHM quiver and circular quiver Chern-Simons (including ABJM) theories with specific boundary conditions. 
We note matching of anomalies and present the half-indices which have been checked to match to high order in $q$, 
and in some cases the matching is checked exactly. 
Here we focus on the case of Abelian theories where Nahm pole boundary conditions are simply Dirichlet boundary conditions.

\begin{figure}
\begin{tikzpicture}
    \node (A) at (0,0) [rectangle, draw] {\begin{tabular}{c} $U(1)-[l]$ ADHM \\ $(p, q)$ \end{tabular}};
    \node (B) at (8.7,0) [rectangle, draw] {\begin{tabular}{c} $U(1)_1 \times U(1)^{l-1} \times U(1)_{-1}$ circular quiver CS \\ $(p+q, q)$ \end{tabular}};
    \node (C) at (0,-3) [rectangle, draw] {\begin{tabular}{c} $\widetilde{U(1)^l}$ ADHM quiver \\ $(q, p)$ \end{tabular}};
    \node (D) at (8.4,-3) [rectangle, draw] {\begin{tabular}{c} $U(1)_1 \times \widetilde{U(1)^{l-1}} \times U(1)_{-1}$ circular quiver CS \\ $(p+q, p)$ \end{tabular}};
	\draw[->,ultra thick] (A) -- node[above] {$-STS$} (B);
    \draw[->,ultra thick] (A) -- node[right] {$\Pcal_{123} S$} (C);
    \draw[->,ultra thick] (C) -- node[above] {$-STS$} (D);
\end{tikzpicture}
\caption{Abelian dualities with mapping of boundary $(p, q)$ 5-branes for the ADHM theories and circular quiver Chern-Simons theories.}
\label{Fig_Abelian_dualities}
\end{figure}
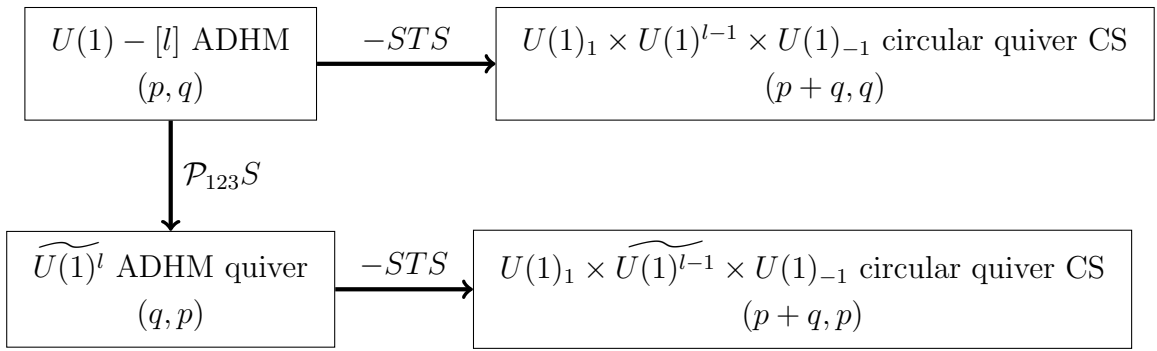

The Abelian dualities for $l \ge 1$ are summarized in  Figure~\ref{Fig_Abelian_dualities}. 
We also note the mapping of $(p, q)$ 5-branes in the ADHM theory under the dualities. 
The labels on the arrows denote the $SL(2, \Zb)$ dualities along with $\Pcal_{123}$ which is a parity transformation reversing the orientation of $x^1$, $x^2$ and $x^3$. In Figure~\ref{Fig_Abelian_dualities_l1} we indicate explicitly the theories in the case of $l = 1$ where the circular quiver becomes the ABJM theory and the ADHM quiver is the mirror ADHM theory.

\begin{figure}
\begin{tikzpicture}
    \node (A) at (0,0) [rectangle, draw] {\begin{tabular}{c} $U(1)-[1]$ ADHM \\ $(p, q)$ \end{tabular}};
    \node (B) at (10,0) [rectangle, draw] {\begin{tabular}{c} $U(1)_1 \times U(1)_{-1}$ ABJM \\ $(p+q, q)$ \end{tabular}};
    \node (C) at (0,-3) [rectangle, draw] {\begin{tabular}{c} $\widetilde{U(1)}-[1] \; \textrm{ADHM}$ \\ $(q, p)$ \end{tabular}};
    \node (D) at (10,-3) [rectangle, draw] {\begin{tabular}{c} $U(1)_1 \times U(1)_{-1} \; \textrm{ABJM}$ \\ $(p+q, p)$ \end{tabular}};
	\draw[->,ultra thick] (A) -- node[above] {$-STS$} (B);
    \draw[->,ultra thick] (A) -- node[right] {$\Pcal_{123} S$} (C);
    \draw[->,ultra thick] (C) -- node[above] {$-STS$} (D);
\end{tikzpicture}
\caption{Abelian dualities with mapping of boundary $(p, q)$ 5-branes for the ADHM and ABJM theories.}
\label{Fig_Abelian_dualities_l1}
\end{figure}
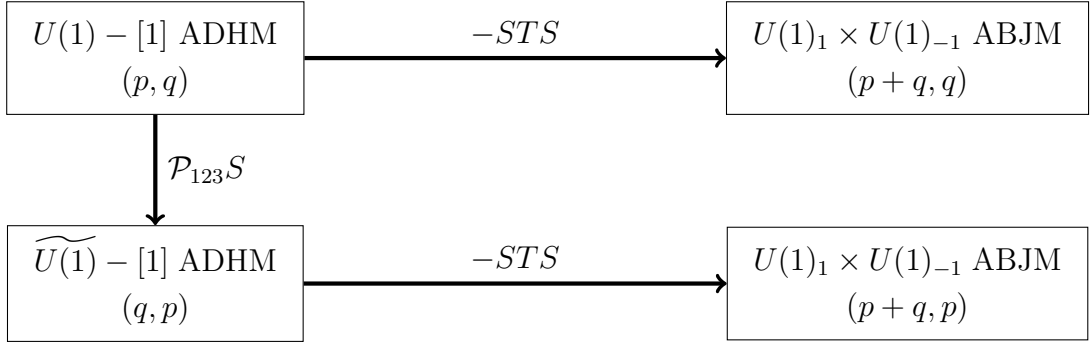

\subsection{$U(1)-[1]$ ADHM $(\mathcal{N},N,N)+\widetilde{\Gamma}+\widetilde{\eta}$ and $(\mathcal{D},D,D)$ dualities}
\label{Sec_ADHM_U1_1_NNN_duals}
Consider the $\mathcal{N}=(0,4)$ boundary condition $(\mathcal{N},N,N)+\widetilde{\Gamma}+\widetilde{\eta}$ in the $U(1)$ ADHM theory with one flavor, 
where the $U(1)$ vector multiplet obeys Neumann boundary condition $\mathcal{N}$, 
the neutral hyper is subject to Neumann boundary condition $N$ 
and the charged hyper satisfies Neumann boundary condition $N$ 
together with the charged Fermi multiplet $\widetilde{\Gamma}$ which cancels the gauge anomaly and the Fermi $\widetilde{\eta}$. The general theory was described in Section~\ref{Sec_ADHM_NNN_Anom} and in this case the anomaly was shown in \eqref{bdy_U1_1_ADHM_anom_NDNN_Gamma_eta} to be
\begin{align}
\Acal = & z^2 + u_1^2 - 2u_1x + 2tr \; .
\end{align}

The bulk ADHM theory is dual to the ABJM theory and we claim this extends to the boundary conditions considered here. In the brane configuration this arises from the $-STS$ duality transformation. Specifically we claim that in the $U(1)_1 \times U(1)_{-1}$ ABJM theory we have the boundary conditions for the vector multiplets as described in section~\ref{Sec_ABJM_U1k_diag}. As explained there, this leads to a description in terms of free chirals and indeed we see that we get anomaly matching with a hypermultiplet charged under $U(1)_x$ with $R$-charge $\frac{1}{2}$ 
having Neumann boundary condition, and a twisted hypermultiplet charged under $U(1)_z$ with $R$-charge $\frac{1}{2}$ 
having Dirichlet boundary condition, together with the 2d Fermi multiplet $\widetilde{\eta}$ charged under $U(1)_u$ and $U(1)_x$, so
\begin{align}
\label{bdy_hyper_thyper_ND_eta}
\Acal = & \frac{1}{2}\left( z - t - \frac{1}{2}r \right)^2 + \frac{1}{2}\left( -z - t - \frac{1}{2}r \right)^2 
\nonumber \\
 & - \frac{1}{2}\left( x + t - \frac{1}{2}r \right)^2 - \frac{1}{2}\left( -x + t - \frac{1}{2}r \right)^2 + (u - x)^2
\nonumber \\
 & = z^2 + u^2 - 2ux + 2tr \; .
\end{align}
We see this matches up to labelling $u \equiv u_1$.

Also, as the bulk ADHM theory is self-mirror, the boundary condition is expected to map to some boundary condition in the same theory under mirror symmetry
\begin{align}
    \label{mirror11}
    t \to -t , \qquad x \leftrightarrow z \; . 
\end{align}
According to mirror symmetry of $\mathcal{N}=(0,4)$ boundary conditions in Abelian gauge theories in \cite{Okazaki:2019bok}, 
the mirror boundary condition is given by Dirichlet boundary condition $\mathcal{D}$ for vector multiplet 
and the Dirichlet boundary conditions $D$ for adjoint and fundamental hypermultiplets. 
This does have the same anomaly as shown from applying this mirror map to \eqref{bdy_U1_1_ADHM_anom_DNDD} and is the expected result from the brane configuration under the $S$ duality transformation. 
Note that this is a special case of the more general duality between $U(1)-[l]$ ADHM and the $U(1)^l$ ADHM quiver theory and indeed the anomaly is given by setting $l = 1$ in \eqref{bdy_U1_l_ADHMquiver_anom_DNDD}.

Recall that with these boundary conditions, the ADHM half-index is written as (\ref{ADHMu1_1_04Neu})
\begin{align}
\mathbb{II}_{\mathcal{N},N,N+\widetilde{\Gamma}+\widetilde{\eta}}^{\textrm{$U(1)$ ADHM}-[1]}
&=(q)_{\infty}(q^{\frac12}t^2;q)_{\infty}
\frac{(q^{\frac{1}{2}} u_1^{\pm} x^{\mp}; q)_{\infty}}{(q^{\frac14}tx^{\pm};q)_{\infty}}
\oint \frac{ds}{2\pi is}
\frac{(q^{\frac12}s^{\pm}z^{\mp};q)_{\infty}}
{(q^{\frac14}ts^{\pm};q)_{\infty}}. 
\end{align}

The half-index for the ABJM theory or equivalently the free chirals, along with the Fermi $\widetilde{\eta}$ is given by the expression
\begin{align}
\label{free_04NeuDir_eta}
\frac{(q^{\frac34}tz^{\pm};q)_{\infty}}{(q^{\frac14}tx^{\pm};q)_{\infty}}
 (q^{\frac12}u_1^{\pm}x^{\mp};q)_{\infty} \; . 
\end{align}

The mirror ADHM half-index is given by applying the mirror map\footnote{For fugacities this is $t \to t^{-1}$ and $x \leftrightarrow z$.} to \eqref{h_ADHMu1nfl_DDD} and setting $l = 1$
\begin{align}
\label{ADHMu1_04Dir}
\mathbb{II}_{\mathcal{D},D,D}^{\widetilde{\textrm{$U(1)$ ADHM}-[1]}}
&=\frac{1}{(q)_{\infty}(q^{\frac12}t^2;q)_{\infty}}
\sum_{m\in \mathbb{Z}}
(-1)^m q^{\frac{m^2}{2}}u_1^m x^{-m} 
(q^{\frac34}tz^{\pm};q)_{\infty}
(q^{\frac34\pm m}tu_1^{\pm};q)_{\infty} \; .
\end{align}

We claim that all three of these half-indices are equal and this has been checked to high order in $q$.

To summarize we conjecture the following dualities
\begin{align}
\label{u1_1_dualities_NNN}
    & U(1)-[1] \textrm{ ADHM with } (\Ncal, N, N) + \widetilde{\Gamma} + \widetilde{\eta} \textrm{ from } \textrm{NS}5'
    \nonumber \\
    \longleftrightarrow & U(1)_1 \times U(1)_{-1} \; \textrm{ABJM \; with \;} (N, D) + \widetilde{\eta} \textrm{ from } \textrm{NS}5'
    \nonumber \\
    \longleftrightarrow & U(1)-[1] \; \widetilde{\textrm{ADHM}} \textrm{\; with \;} (\Dcal, D, D) \textrm{ from } \textrm{D}5'\; .
\end{align}

Note that the Fermi $\widetilde{\eta}$ could be removed from the ADHM and ABJM or free chiral theories but it is required for the ADHM mirror duality unless we instead introduce a 2d chiral multiplet in the mirror ADHM theory with Dirichlet boundary conditions. In any case we choose to keep the Fermi $\widetilde{\eta}$ as it is expected from the brane configuration.

\subsection{$U(1)-[1]$ ADHM $(\mathcal{N},D,N)+\widetilde{\Gamma}$ duality}
\label{Sec_ADHM_U1_1_NDN_duals}

We consider here boundary conditions for the ADHM theory which are straightforward for the Abelian theory. In particular, we take the $\mathcal{N}=(0,4)$ boundary condition $(\mathcal{N},D,N)+\Gamma$ in the $U(1)$ ADHM theory with one flavor, 
where the $U(1)$ vector multiplet obeys Neumann boundary condition $\mathcal{N}$, 
the neutral hyper is subject to Dirichlet boundary condition $D$ 
and the charged hyper satisfies Neumann boundary condition $N$ 
together with the charged Fermi multiplet $\widetilde{\Gamma}$ which cancels the gauge anomaly.
We can easily calculate the gauge and 't Hooft anomalies as follows,
\begin{align}
\label{bdy_U1_1_ADHM_anom_NDN}
\Acal & = \underbrace{\frac{1}{2}r^2}_{\textrm{VM}, \; \Ncal} - \underbrace{2sz}_{\textrm{FI}} + \underbrace{2t^2}_{\Phi, \; D}
 + \underbrace{\left( x^2 + \left( t - \frac{1}{2}r \right)^2 \right)}_{(X, Y), \; D}
- \underbrace{\left( s^2 + \left( t - \frac{1}{2}r \right)^2 \right)}_{(I, J), \; N}
  + \underbrace{\Tr(s + z)^2}_{\widetilde{\Gamma}}
  \nonumber \\
  & = x^2 + 2t^2 + \frac{1}{2}r^2 + z^2 \; .
\end{align}

The half-index takes the form
\begin{align}
\label{h_u1nf1_NDN}
&
\mathbb{II}_{\mathcal{N},D,N+\widetilde{\Gamma}}^{\textrm{$U(1)$ ADHM}-[1]}
\nonumber\\
&=
(q)_{\infty} (q^{\frac12}t^2;q)_{\infty}
\oint \frac{ds}{2\pi is}
(q^{\frac34}t^{-1}x^{\pm};q)_{\infty}
\frac{1}{(q^{\frac14}ts^{\pm};q)_{\infty}}
(q^{\frac12} s^{\pm}z^{\mp};q)_{\infty}. 
\end{align}
We find that the half-index (\ref{h_u1nf1_NDN}) agrees with the half-index 
\begin{align}
\label{h_HM+tHM_DD}
\mathbb{II}_{D}^{\textrm{HM}}(x)\mathbb{II}_{D}^{\textrm{tHM}}(z)
&=(q^{\frac34}t^{-1}x^{\pm};q)_{\infty}(q^{\frac34}tz^{\pm};q)_{\infty} 
\end{align}
of the free theory with hypermultiplet obeying the Dirichlet boundary condition $D$ 
and a free twisted hypermultiplet satisfying the Dirichlet boundary condition $D$.
This arises from $U(1)_1 \times U(1)_{-1}$ ABJM if we take Dirichlet boundary conditions for the hypermultiplet $(H, \widetilde{H})$ and Dirichlet boundary conditions for the twisted hypermultiplet $(T, \tilde{T})$ and if we have no contribution from the vector multiplet or Chern-Simons levels, which can be explained following the arguments in Section~\ref{Sec_ABJM_U1k_diag}.
In particular, we have matching anomaly \eqref{bdy_U1_ABJM_HtH_anom_ND} when we incorporate the diagonal breaking of the gauge group so $s_1 = s_2$.

Interestingly, we see that the half-indices \eqref{h_u1nf1_NDN} and \eqref{h_HM+tHM_DD} are invariant under the mirror transformation \eqref{mirror11}. 
This is manifest for the anomaly and for the $(D,D)$ boundary condition in the free hypermultiplet and twisted hypermultiplet half-index \eqref{h_HM+tHM_DD}. This means that the boundary condition $(\mathcal{N},D,N)+ \widetilde{\Gamma}$ in the $U(1)-[1]$ ADHM theory is dual to the $(\Ncal, D, N) + \Gamma$ boundary condition in the mirror $U(1)-[1]$ $\widetilde{\textrm{ADHM}}$ theory since we can easily see that \eqref{h_u1nf1_NDN} maps to
\begin{align}
\label{h_u1nf1_NDN_mirror}
&
\mathbb{II}_{\mathcal{N},D,N+\Gamma}^{U(1)-[1] \widetilde{\textrm{ADHM}}}
\nonumber\\
&=
(q)_{\infty} (q^{\frac12}t^{-2};q)_{\infty}
\oint \frac{ds}{2\pi is}
(q^{\frac34}t z^{\pm};q)_{\infty}
\frac{1}{(q^{\frac14}t^{-1} s^{\pm};q)_{\infty}}
(q^{\frac12} s^{\pm} x^{\mp};q)_{\infty}. 
\end{align}
The Fermi $\widetilde{\Gamma}$ in the ADHM theory and the Fermi $\Gamma$ in the $\widetilde{\textrm{ADHM}}$ theory are the same except for the exchange of $U(1)_z$ charge with $U(1)_x$ charge.

To summarize we conjecture the following dualities
\begin{align}
    & U(1)-[1] \textrm{ ADHM with } (\Ncal, D, N) + \widetilde{\Gamma} \textrm{ from } (-1, 1)'  \; 5\textrm{-brane}
    \nonumber \\
    \longleftrightarrow & U(1)_1 \times U(1)_{-1} \textrm{ ABJM with } (D, D) \textrm{ from D}5'\textrm{-brane}
    \nonumber \\
    \longleftrightarrow & U(1)-[1] \; \widetilde{\textrm{ADHM}} \textrm{ with } (\Ncal, D, N) + \Gamma \textrm{ from } \widetilde{(1, -1)}'  \; 5\textrm{-brane} \; .
\end{align}
We explain the identification of the boundary $(p, q)$ 5-branes below.

If we take the case of $U(N)-[1]$ ADHM then S-duality maps the brane configuration
\eqref{IIB_branesetup_ADHM} to \eqref{IIB_branesetup_ADHMquiver}
which is equivalent -- it can be seen by relabelling the $345$ and $789$ directions that we arrive back at the original brane configuration. Hence we see that the bulk $U(N)-[1]$ ADHM theory has a self-mirror property.

However, if we introduce one of the boundary branes NS5$'$ or D5$'$ we see that they are not invariant and in fact are mapped to each other under this mirror transformation. Hence we expect to have a mirror map of boundary conditions in the $U(N)-[1]$ ADHM theory relating Neumann to Nahm pole for the Vector Multiplet. In the Abelian case this is simply Neumann to Dirichlet as we have seen in Section~\ref{Sec_ADHM_U1_1_NNN_duals}.

It is interesting to ask what the effect of more general $(p, q)$ 5-branes as boundary branes would be and this is not immediately clear. Note that there are two possible types corresponding to bound states of the branes we have labelled D5 and NS5$'$ or of NS5 and D5$'$. In both cases the $(p,q)$ 5-branes would be oriented at a specific angle in the $26$-plane. However, they would differ in whether they filled the $345$ or the $789$ space. So, for a given choice of $(p, q)$ we have two different boundary 5-branes, except for $(1, 0)$ and $(0, 1)$ where one of the two choices gives a brane in the bulk ADHM configuration so does not provide a boundary. Since the D3-branes do not distinguish between the $345$ and $789$ directions we conjecture that the two different types of $(p, q)$ 5-branes would provide essentially identical boundary conditions in the field theory but different 2d Fermis would arise from the 5-brane intersections. Therefore, since the mirror transformation (S-duality including the parity transformation described after \eqref{IIB_branesetup_ADHMquiver} along with relabelling the $345$ and $789$ directions) maps a $(p, q)$ 5-brane of one type to a $(q, p)$ 5-brane of the other type, we may have self-mirror boundary conditions in the case of a $(1, 1)$ 5-brane. There is also the possibility that different boundary $(p, q)$ 5-branes give the same field theory boundary conditions. We conjecture that this occurs only for the pair $(p, q)$ and $(-p, -q)$, so the other case expected to lead to self-mirror boudnary conditions in the ADHM theory is a boundary $(-1, 1)$ or $(1, -1)$ 5-brane. This is consistent from the field theory matching of anomalies and half-indices we saw in Section~\ref{Sec_ADHM_U1_1_NDN_duals} with the different, but very similar, Fermis arising from the different $(-1, 1)$ or $(1, -1)$ 5-brane orientations.

In claiming that the boundary conditions in Section~\ref{Sec_ADHM_U1_1_NDN_duals} arise from $(-1, 1)'$ 5-brane in the ADHM theory, rather than a $(1, 1)'$ 5-brane we note that the former results in a boundary D5'-brane in the ABJM theory. Here the brane configuration includes the NS5-brane with the bi-fundamental hypermultiplet $(H, \widetilde{H})$ arising from the fundamental strings with ends on the D3-branes on both sides on the NS5-brane. It is well known that a boundary D5'-brane would provide Dirichlet boundary conditions for this hypermultiplet, consistent with the boundary conditions described. As we will see next, the alternative self-mirror boundary condition has Neumann boundary conditions for the hypermultiplet so must arise from the other choice of boundary 5-brane.

\subsection{$U(1)-[1]$ ADHM $(\mathcal{D},N,D)$ duality}
We can consider the $\mathcal{N}=(0,4)$ boundary condition $(\mathcal{D},N,D)$ in the $U(1)$ ADHM theory with one flavor, 
where the vector multiplet satisfies Dirichlet boundary condition $\mathcal{D}$, 
the neutral hyper obeys Neumann boundary condition $N$ 
and the charged hyper has the Dirichlet boundary condition $D$.
We can easily calculate the gauge and 't Hooft anomalies as follows,
\begin{align}
\label{bdy_U1_1_ADHM_anom_DND}
\Acal = & - \underbrace{\frac{1}{2}r^2}_{\textrm{VM}, \; \Dcal} - \underbrace{2uz}_{\textrm{FI}} - \underbrace{2t^2}_{\Phi, \; N}
 - \underbrace{\left( x^2 + \left( t - \frac{1}{2}r \right)^2 \right)}_{(X, Y), \; N}
 + \underbrace{\left( u^2 + \left( t - \frac{1}{2}r \right)^2 \right)}_{(I, J), \; D}
  \nonumber \\
  = & u^2 - 2uz - x^2 - 2t^2 - \frac{1}{2}r^2 \; .
\end{align}
The anomaly is matched by free chirals or $U(1)_1 \times U(1)_{-1}$ ABJM with Neumann boundary conditions for the hypermultiplet and twisted hypermultiplet along with a Fermi multiplet $\Lambda$ with $U(1)_u$ charge $+1$ and $U(1)_z$ charge $-1$,
\begin{align}
\label{bdy_HM_tHM_Fermi_NN}
\Acal = & - \underbrace{\left( x^2 + \left( t - \frac{1}{2}r \right)^2 \right)}_{\textrm{HM}, \; N}
 - \underbrace{\left( z^2 + \left( t + \frac{1}{2}r \right)^2 \right)}_{\textrm{tHM}, \; N}
 + \underbrace{\left( u - z \right)^2}_{\textrm{Fermi $\Lambda$}}
  \nonumber \\
  = & u^2 - 2uz - x^2 - 2t^2 - \frac{1}{2}r^2 \; .
\end{align}

The matching half-indices are
\begin{align}
\label{h_ADHMu1nf1_DND}
\mathbb{II}_{\mathcal{D},N,D}^{\textrm{$U(1)$ ADHM}-[1]}
&=\frac{1}{(q)_{\infty} (q^{\frac12}t^{-2};q)_{\infty}}
\sum_{m\in \mathbb{Z}}
(-1)^m q^{\frac{m^2}{2}} u^m z^m
\frac{1}
{(q^{\frac14}tx^{\pm};q)_{\infty}}
(q^{\frac34\pm m} t^{-1}u^{\pm};q)_{\infty} 
\end{align}
and
\begin{align}
\label{h_HM+tHM_NN_Fermi}
\mathbb{II}_{N}^{\textrm{HM}}(x)
\mathbb{II}_{N}^{\textrm{tHM}}(z)
F(q^{\frac12} u z^{-1}) \; .
\end{align}

There is no obvious set of mirror ADHM boundary conditions but if we introduce a 2d Fermi $\Gamma$ with $U(1)_u$ charge $+1$ and $U(1)_x$ charge $-1$ then we see that we have self-mirror configurations with
\begin{align}
\label{h_ADHMu1nf1_DND_duals}
\mathbb{II}_{(\Dcal,N,D) + \Gamma}^{U(1) \textrm{ADHM}-[1]} & =
\mathbb{II}_{(\Dcal,N,D) + \widetilde{\Gamma}}^{U(1) \widetilde{\textrm{ADHM}}-[1]} =
\mathbb{II}_{N}^{\textrm{HM}}(x)
\mathbb{II}_{N}^{\textrm{tHM}}(z)
F(q^{\frac12} u x^{-1}) F(q^{\frac12} u z^{-1}) \; .
\end{align}

To summarize we conjecture the following dualities
\begin{align}
    & U(1)-[1] \textrm{\; ADHM \; with \;} (\Dcal, N, D) + \Gamma
    \textrm{ from } (1, 1)'  \; 5\textrm{-brane}
    \nonumber \\
    \longleftrightarrow & U(1)_1 \times U(1)_{-1} \; \textrm{ABJM \; with \;} (N, N) + \Lambda + \Gamma
    \textrm{ from } (2, 1)'  \; 5\textrm{-brane}
   \nonumber \\
    \longleftrightarrow & U(1)-[1] \; \widetilde{\textrm{ADHM}} \textrm{\; with \;} (\Dcal, N, D) + \widetilde{\Gamma}
    \textrm{ from } \widetilde{(1, 1)}'  \; 5\textrm{-brane} \; .
\end{align}

Here we have conjectured the boundary branes. 
We previously noted that the $(-1,1)'$ 5-brane was expected to provide the self-mirror boundary conditions for the ADHM theory in Section~\ref{Sec_ADHM_U1_1_NDN_duals}. 
Here we have another set of self-mirror boundary conditions and this time we argue that this is provided by the $(1,1)'$ 5-brane which maps to a $\widetilde{(1,1)}'$ 5-brane. 
This then leads to the conjecture that the $(2,1)'$ 5-brane provides the Nuemann boundary conditions for the hypermultiplet and twisted hypermultiplet in the ABJM theory.


Finally, we note that there is another self-mirror $\mathcal{N}=(0,4)$ boundary condition $(\mathcal{D},N,\tilde{D})$ in the $U(1)$ ADHM theory with one flavor, arising from the $(\mathcal{D},N,D)$ boundary condition where the charged hypermultiplet has the Dirichlet boundary condition $\tilde{D}$ deformed by a boundary vev, 
corresponding to the specialization of the fugacity $u$ for the boundary global symmetry resulting from the gauge group as $u=z$.
We can easily calculate the gauge and 't Hooft anomalies from the $(\mathcal{D},N,D)$ boundary condition case (without introducing the Fermi $\Lambda$) by setting the field strength $u = z$ in \eqref{bdy_U1_1_ADHM_anom_DND} which gives
\begin{align}
\label{bdy_U1_1_ADHM_anom_DNDc}
\Acal = & - z^2 - x^2 - 2t^2 - \frac{1}{2}r^2 \; .
\end{align}
The half-index is
\begin{align}
\label{h_ADHMu1nf1_DNDc}
\mathbb{II}_{\mathcal{D},N,\tilde{D}}^{\textrm{$U(1)$ ADHM}-[1]}
&=\frac{1}{(q)_{\infty} (q^{\frac12}t^{-2};q)_{\infty}}
\sum_{m\in \mathbb{Z}}
(-1)^m q^{\frac{m^2}{2}}
\frac{1}
{(q^{\frac14}tx^{\pm};q)_{\infty}}
(q^{\frac34\pm m} t^{-1}z^{\pm};q)_{\infty}, 
\end{align}
which matches the half-index
\begin{align}
\label{h_HM+tHM_NN_Fermi_c}
\mathbb{II}_{N}^{\textrm{HM}}(x)
\mathbb{II}_{N}^{\textrm{tHM}}(z)
F(q^{\frac12}) \; .
\end{align}
We note that the half-indices (\ref{h_ADHMu1nf1_DNDc}) and (\ref{h_HM+tHM_NN_Fermi_c}) are invariant under (\ref{mirror11}). 
This implies that the boundary condition $(\mathcal{D},N,\tilde{D})$ is self-mirror.

\subsection{$U(1)-[l>1]$ ADHM $(\mathcal{N},N,N)+\widetilde{\Gamma}+\widetilde{\eta}$ duality}

From dualities of the brane configurations we expect $U(1)-[l]$ ADHM theory with boundary conditions $(\Ncal, N, N)$ along with Fermis $\widetilde{\Gamma}_{I, I+1}$ and $\widetilde{\eta}$ to be dual to the $U(1)^l$ ADHM quiver theory with $(\Dcal, D, D)$ boundary conditions and to the $U(1)_1 \times U(1)^{l-1} \times U(1)_{-1}$ circular quiver Chern-Simons theory with $(\Dcal, N, D)$ boundary conditions and Fermi $\eta$. Indeed, for simplicity ignoring the $U(1)_{y_{\alpha}}$ in this section although they can be straightforwardly included, the anomalies \eqref{bdy_U1_l_ADHM_anom_NDNN_Gamma}, \eqref{bdy_U1_l_ADHMquiver_anom_DNDD} and \eqref{bdy_U1_lp1_CQCS_anom_DND_eta} match with evaluation
\begin{align}
\label{bdy_U1_l_ADHM_anom_NDNN_Gamma_duals}
\Acal = & \sum_{I=1}^l \left( u_I - u_{I+1} \right)^2 + lz^2 + u_1^2 - 2 u_1 x
  \nonumber \\
  &  + (1-l)t^2 + (1+l)tr + \frac{1 - l}{4}r^2. 
\end{align}


Starting with the Neumann ADHM half-index \eqref{ADHMu1_l_04Neu}
\begin{align}
&
\mathbb{II}_{\mathcal{N},N,N+\widetilde{\Gamma}+\widetilde{\eta}}^{\textrm{$U(1)$ ADHM}-[l]}(t,x,z,u_I;q)
\nonumber\\
& = (q)_{\infty} (q^{\frac12}t^2;q)_{\infty}
\oint \frac{ds}{2\pi is}
 \frac{(q^{\frac{1}{2}} u_1^{\pm} x^{\mp}; q)_{\infty}}{(q^{\frac14}tx^{\pm};q)_{\infty}}
\frac{\prod_{I = 1}^l (q^{\frac12}s^{\pm} u_I^{\pm} u_{I+1}^{\mp} z^{\pm};q)_{\infty}}
{(q^{\frac14}ts^{\pm};q)_{\infty}^l}, 
\end{align}
the minimal mirror identity \cite{Gaiotto:2019jvo}
\begin{align}
    \label{eq_min_mirror_id}
    \frac{(q^{\frac12}s^{\pm} x^{\pm};q)_{\infty}}{(q^{\frac14}ts^{\pm};q)_{\infty}} & = \frac{1}{(q)_{\infty} (q^{\frac{1}{2}} t^2; q)_{\infty}} \sum_{m \in \Zb} q^{\frac{1}{2} m^2} (-sx)^m
 (q^{\frac{3}{4} \pm m} t x^{\pm}; q)_{\infty} \; 
\end{align}
can be used to write the contribution from the $l$ fundamentals $(I_{\alpha}, J_{\alpha})$ and Fermis $\widetilde{\Gamma}_{I, I+1}$ as $l$ sums
\begin{align}
&\prod_{I = 1}^l \frac{(q^{\frac12}s^{\pm} u_I^{\pm} u_{I+1}^{\mp} z^{\pm};q)_{\infty}}
{(q^{\frac14}ts^{\pm};q)_{\infty}}
\nonumber\\
&= \frac{1}{(q)_{\infty}^l (q^{\frac{1}{2}} t^2; q)_{\infty}^l} \sum_{m_I \in \Zb} q^{\frac{1}{2}\sum_{I=1}^l m_I^2} (-sz)^{\sum_{I=1}^l m_I} \prod_{I=1}^l u_I^{m_I} u_{I+1}^{-m_I}
 (q^{\frac{3}{4} \pm m_I} t u_I^{\pm} u_{I+1}^{\mp} z^{\pm}; q)_{\infty} \; .
\end{align}
The contour integral over $s$ then fixes $\sum_{I = 1}^l m_I = 0$ resulting in a rewriting of the half-index as
\begin{align}
\label{ADHMu1_04Neu_MinMirrorID}
&
\mathbb{II}_{\mathcal{N},N,N+\widetilde{\Gamma}+\widetilde{\eta}}^{\textrm{$U(1)$ ADHM}-[l]}(t,x,z,u_I;q)
\nonumber\\
&= \frac{1}{(q)_{\infty}^{l-1} (q^{\frac12}t^2;q)_{\infty}^{l-1}}
\frac{(q^{\frac{1}{2}} u_1^{\pm} x^{\mp}; q)_{\infty}}{(q^{\frac14}tx^{\pm};q)_{\infty}}
\sum_{m_I \in \Zb \vert \sum_I m_I = 0} q^{\frac{1}{2}\sum_I m_I^2} \prod_I u_I^{m_I} u_{I+1}^{-m_I}
 (q^{\frac{3}{4} \pm m_I} t u_I^{\pm} u_{I+1}^{\mp} z^{\pm}; q)_{\infty}
 \nonumber \\
 & = 
\mathbb{II}_{(\Dcal,N,D)+\eta}^{\textrm{$U(1)^{l+1}$ Quiver CS}}(t,x,z,u_I;q), 
\end{align}
where we see that we have derived the expression \eqref{QuiverCS_04Dir_int_red} for the circular quiver Chern-Simons theory with $(\Dcal,N,D)+\eta$ boundary conditions, 
noting the identification $v_I = u_I u_{I+1}^{-1}$.
Note that although we have set the fugacities $y_{\alpha} = 1$ throughout it is straightforward to include them and the matching of half-indices can easily be checked analytically in the general case using the minimal mirror identity.

We claim that these half-indices also match the Dirichlet $U(1)^l$ ADHM quiver half-index \eqref{h_ADHMquiver_u1_l_DDD}.


The case of $l = 1$ has already been presented where the circular quiver Chern-Simons theory is the ABJM theory. 
For $l = 2$ the conjectured matching of half-indices is to \eqref{h_ADHMquiver_u1_2_DDD}
\begin{align}
& \mathbb{II}_{\mathcal{N},N,N+\widetilde{\Gamma}+\widetilde{\eta}}^{\textrm{$U(1)$ ADHM}-[2]}(t,x,z,u_I;q)
\nonumber\\
= & (q)_{\infty} (q^{\frac12}t^2;q)_{\infty}
\oint \frac{ds}{2\pi is}
\times \frac{(q^{\frac{1}{2}} u_1^{\pm} x^{\mp}; q)_{\infty}}{(q^{\frac14}tx^{\pm};q)_{\infty}}
\frac{\prod_{I = 1}^2 (q^{\frac12}s^{\pm} u_I^{\pm} u_{I+1}^{\mp} z^{\pm};q)_{\infty}}
{(q^{\frac14}ts^{\pm};q)_{\infty}^2}
\nonumber \\
= & \mathbb{II}_{(\Dcal,N,D)+\eta}^{\textrm{$U(1)^{3}$ Quiver CS}}(t,x,z,u_I;q)
\nonumber \\
= & \frac{1}{(q)_{\infty} (q^{\frac12}t^2;q)_{\infty}}
\frac{(q^{\frac{1}{2}} u_1^{\pm} x^{\mp}; q)_{\infty}}{(q^{\frac14}tx^{\pm};q)_{\infty}}
\sum_{m_I \in \Zb \vert \sum_I m_I = 0} q^{\frac{1}{2}\sum_I m_I^2} \prod_I u_I^{m_I} u_{I+1}^{-m_I}
 (q^{\frac{3}{4} \pm m_I} t u_I^{\pm} u_{I+1}^{\mp} z^{\pm}; q)_{\infty}
\nonumber \\
= &
\mathbb{II}_{\mathcal{D},D,D}^{\textrm{$U(1)^2$ ADHM quiver}}
\nonumber \\
= & \frac{1}{(q)_{\infty}^2 (q^{\frac12}t^2;q)_{\infty}^2} \sum_{m_I \in \Zb} (-1)^{m_1} q^{\frac{3}{2} m_1^2 + m_2^2 - 2 m_1 m_2} x^{-m_1}
\nonumber \\
 & 
\times (q^{\frac34 \pm m_1 \mp m_2} t u_1^{\pm} u_2^{\mp} z^{\mp};q)_{\infty}
(q^{\frac34 \pm m_1 \mp m_2} t u_1^{\pm} u_2^{\mp} z^{\pm};q)_{\infty}
u_1^{3m_1 - 2m_2} u_2^{2m_2 - 2m_1}
(q^{\frac34 \pm m_1}t u_1^{\pm};q)_{\infty} \; .
\end{align}

To summarize we conjecture the following dualities
\begin{align}
\label{u1_l_dualities_NNN}
    & U(1)-[l] \textrm{ ADHM with } (\Ncal, N, N) + \widetilde{\Gamma} + \widetilde{\eta} \textrm{ from NS}5'
    \nonumber \\
    \longleftrightarrow & U(1)_1 \times U(1)^{l-1} \times U(1)_{-1} \; \textrm{circular quiver CS with } (\Dcal, N, D) + \eta \textrm{ from NS}5'
    \nonumber \\
    \longleftrightarrow & U(1)^l \textrm{ ADHM quiver with } (\Dcal, D, D) \textrm{ from D}5'\; .
\end{align}

\subsection{$U(1)-[l>1]$ ADHM $(\Dcal,D,D)$ duality}
Now taking Dirichlet boundary conditions in the $U(1)-[l]$ ADHM theory we see that the anomaly \eqref{bdy_U1_l_ADHM_anom_DDD} matches the Neumann $U(1)^l$ ADHM quiver theory with Fermis $\Gamma$ and $\eta$ \eqref{bdy_U1_l_ADHMquiver_anom_NDNN_Fermi} as well as the $U(1)_1 \times U(1)^{l-1} \times U(1)_{-1}$ circular quiver Chern-Simons theory with $(\Ncal, D, N)$ boundary conditions and Fermis $\eta_{I}$ \eqref{bdy_U1_lp1_CQCS_anom_NDN_eta_l}
\begin{align}
\label{bdy_U1_l_ADHM_anom_DDD_duals}
\Acal = & l u^2 - 2lzu + x^2 + (l-1)t^2 - (l+1)tr + \frac{l-1}{4}r^2 \; ,
\end{align}
where again for simplicity we will ignore the $U(1)_{y_{\alpha}}$ in this section.

We claim the half-indices \eqref{h_ADHMu1nfl_DDD}, \eqref{ADHMquiveru1_04Neu} and \eqref{QuiverCS_04Neu_int_red} match, i.e.
\begin{align}
\label{h_ADHMu1nfl_DDD_duals}
&\mathbb{II}_{\mathcal{D},D,D}^{\textrm{$U(1)$ ADHM}-[l]}
= \frac{(q^{\frac{3}{4}} t^{-1} x^{\pm}; q)_{\infty}}{(q)_{\infty} (q^{\frac12}t^{-2};q)_{\infty}} \sum_{m \in \Zb} (-1)^{l m} q^{\frac{l}{2} m^2} z^{-l m}
(q^{\frac34 \pm m}t^{-1}u^{\pm};q)_{\infty}^l
u^{l m}
\nonumber \\
=&\mathbb{II}_{(\Ncal,D,N)+\eta_I}^{\textrm{$U(1)^{l+1}$ Quiver CS}}(t,x,z,u_I;q)
= (q)_{\infty}^{l-1} (q^{\frac12}t^{-2};q)_{\infty}^{l-1}(q^{\frac34} t^{-1} x^{\pm};q)_{\infty}
\nonumber \\
&\times \left( \prod_{I = 1}^{l-1} \frac{ds_I}{2\pi i s_I} \right)
\prod_{I = 1}^l \frac{(q^{\frac{1}{2}} s_I^{\pm} u^{\pm} z^{\mp}; q)_{\infty}}{(q^{\frac{1}{4}} s_I^{\pm} z^{\pm} t^{-1}; q)_{\infty}}
\nonumber \\
=&\mathbb{II}_{\mathcal{N},N,N+\Gamma+\eta}^{\textrm{$U(1)^l$ ADHM quiver}}(t;q)
= (q)_{\infty}^{l} (q^{\frac12}t^{-2};q)_{\infty}^{l} 
\prod_{I = 1}^l \oint \frac{ds^{(I)}}{2\pi is^{(I)}}
\frac{1}
{(q^{\frac14}t^{-1}s^{(I) \pm}s^{(I+1) \mp}z^{\mp};q)_{\infty}}
\nonumber\\
 &\times (q^{\frac12} s^{(I) \pm} s^{(I+1) \mp} u^{\pm} z^{\mp};q)_{\infty}
\frac{1}
{(q^{\frac14}t^{-1}s^{(1) \pm};q)_{\infty}}
(q^{\frac12}s^{(1) \pm}x^{\pm};q)_{\infty}, 
\end{align}
where $s_l = \prod_{I = 1}^{l-1} s_I^{-1}$ in the quiver CS half-index.

To summarize we conjecture the following dualities of boundary conditions
\begin{align}
\label{u1_1_dualities_DDD}
    & U(1)-[l] \textrm{ ADHM with } (\Dcal, D, D) \textrm{ from D}5'
    \nonumber \\
    \longleftrightarrow & U(1)_1 \times U(1)^{l-1} \times U(1)_{-1} \; \textrm{circular quiver CS with } (\Ncal, D, N) + \eta_I \textrm{ from }(1,1)' \; 5\textrm{-brane}
    \nonumber \\
    \longleftrightarrow & U(1)^l \textrm{ ADHM quiver with } (\Ncal, N, N) + \Gamma + \eta \textrm{ from NS}5'\; .
\end{align}

For example, taking $l = 2$, we have
\begin{align}
\label{h_ADHMu1nf2_DDD_duals}
&\mathbb{II}_{\mathcal{D},D,D}^{\textrm{$U(1)$ ADHM}-[2]}
 = \frac{(q^{\frac{3}{4}} t^{-1} x^{\pm}; q)_{\infty}}{(q)_{\infty} (q^{\frac12}t^{-2};q)_{\infty}} \sum_{m \in \Zb} q^{m^2} z^{-2m}
(q^{\frac34 \pm m}t^{-1}u^{\pm};q)_{\infty}^2
u^{2m}
\nonumber \\
= &\mathbb{II}_{(\Ncal,D,N)+\eta_I}^{\textrm{$U(1)^3$ Quiver CS}}(t,x,z,u_I;q)
 =(q)_{\infty} (q^{\frac12}t^{-2};q)_{\infty}
(q^{\frac34} t^{-1} x^{\pm};q)_{\infty}
  \nonumber \\
 &\times \oint \frac{ds}{2\pi i s}
 \frac{(q^{\frac{1}{2}} s^{\pm} u^{\pm} z^{\mp}; q)_{\infty} (q^{\frac{1}{2}} s^{\mp} u^{\pm} z^{\mp}; q)_{\infty}}{(q^{\frac{1}{4}} s^{\pm} z^{\pm} t^{-1}; q)_{\infty} (q^{\frac{1}{4}} s^{\mp} z^{\pm} t^{-1}; q)_{\infty}}
 \nonumber \\
=& \mathbb{II}_{\mathcal{N},N,N+\Gamma+\eta}^{\textrm{$U(1)^2$ ADHM quiver}}(t;q)
 = (q)_{\infty}^2 (q^{\frac12}t^{-2};q)_{\infty}^2
\prod_{I = 1}^2 \oint \frac{ds^{(I)}}{2\pi is^{(I)}}
\frac{1}
{(q^{\frac14}t^{-1}s^{(I) \pm}s^{(I+1) \mp}z^{\mp};q)_{\infty}}
\nonumber\\
 &\times (q^{\frac12} s^{(I) \pm} s^{(I+1) \mp} u^{\pm} z^{\mp};q)_{\infty}
\frac{1}
{(q^{\frac14}t^{-1}s^{(1) \pm};q)_{\infty}}
(q^{\frac12}s^{(1) \pm}x^{\pm};q)_{\infty} \; .
\end{align}

\subsection*{Acknowledgements}
The work of TO was supported by the Startup Funding no.\ 4007012317 of the Southeast University. 
The work of DJS was supported in part by the STFC Consolidated grant ST/T000708/1.

\appendix
\section{Neumann and Dirichlet b.c. in ABJM theory}
\label{ABJM_U1k_VMbc}
Here we systematically consider the cases where each $U(1)_{\pm k}$ vector multiplet has either Neumann or Dirichlet boundary condition and wlog.\ we assume $k > 0$. We further consider only cases where we do not have any 2d chirals charged under a $U(1)_{\pm k}$ for which the vector multiplet has Neumann boundary condition, and we assume no 2d chiral or Fermi multiplets charged under a $U(1)_{\pm k}$ for which the vector multiplet has Dirichlet boundary condition. These conditions mean that in the anomaly polynomial if we have a term $\lambda s^2$ where $s$ is a $U(1)$ field strength then if $\lambda > 0$ we must have Dirichlet boundary condition for that vector multiplet while if $\lambda \le 0$ we must have Neumann boundary condition and if $\lambda < 0$ we must also have 2d Fermi(s) to cancel that gauge anomaly. Let's now focus on the pure gauge anomaly.

In the case of Abelian ABJM, the vector multiplets give no contribution to the gauge anomaly. We therefore only have contributions $k(s_1^2 - s_2^2)$ from the Chern-Simons levels and $\pm \frac{1}{2}(s_1 - s_2)^2$ from each of the 3d chirals $H$, $\tilde{H}$, $T$, $\tilde{T}$ where the minus sign is for Neumann boundary condition and the plus sign for Dirichlet boundary condition. If these chirals have $d$ Dirichlet boundary conditions and $4-d$ Neumann boundary conditions the pure gauge part of the anomaly is
\begin{align}
    \Acal_{\textrm{gauge}, \; d} & = (k - 2 + d)s_1^2 + (2 - d)s_1s_2 - (k + 2 - d)s_2^2 \; .
\end{align}
This means that we need Dirichlet boundary condition for the $U(1)_k$ vector multiplet except in the cases where $k = 2$ and $d = 0$ or where $k = 1$ and either $d = 0$ or $d = 1$ where we must have Neumann boundary condition. For Neumann boundary condition we need a 2d Fermi in the cases of $k = 1$ and $d = 0$. Similarly for the $U(1)_{-k}$ vector multiplet we must have Neumann boundary condition except for the case of $k = 1$ and $d = 4$ where we must have Dirichlet boundary condition. In the cases of Neumann boundary condition we must include 2d Fermi(s) except for the cases of $k = 2$ and $d = 4$ or $k = 1$ and $d = 3$. 
These cases can be summarized in the following table showing which values of $k$ are allowed for each type of boundary condition
\begin{equation}
\begin{array}{c|cccc}
     & \Ncal \Ncal & \Ncal \Dcal & \Dcal \Ncal & \Dcal \Dcal \\ \hline
    d = 0 & 1, (2) &  & \ge 3 &  \\
    d = 1 & (1) &  & \ge 2 &  \\
    d = 2 &  &  & \ge 1 &  \\
    d = 3 &  &  & \ge 1 &  \\
    d = 4 &  &  & \ge 2 & 1 \\
\end{array}
\end{equation}
noting that for $d=1$, $d=2$ and $d=3$ there are options regarding which of the 3d chiral multiplets has each type of boundary condition.
We have highlighted the cases of $d = 0$ and $k = 2$ or $d=1$ and $k=1$ since then we have $\Ncal \Ncal$ boundary conditions for the vector multiplets, but assuming integer gauge charges for 2d Fermis we cannot cancel the mixed anomaly contribution $s_1 s_2$ for $d$ odd, or without introducing 2d chirals for $d = 0$ and $k = 2$. In the case of $d = 1$ or $d = 3$ with $\Dcal \Ncal$ boundary conditions we also cannot cancel the mixed anomaly contribution $(2-d)u_1 s_2$ 
but we can now interpret this as completely breaking the $U(1)_k$ gauge group 
which means we should set the $U(1)_k$ field strength $u_1$ to zero in the anomaly polynomial 
while the $u_!$ fugacity in the half-indices should be set to one and we don't sum over monopole fluxes.
For example, for the boundary condition $(\mathcal{N},\mathcal{N})$ of the vector multiplets in the $U(1)_1\times U(1)_{-1}$ ABJM theory, 
we have the unflavored half-index
\begin{align}
    \II^{d = 0, k = 1}_{\Ncal \Ncal + \Gamma_{12} + 2 \Gamma_2} = & (q)_{\infty}^2 \oint \frac{ds_1}{2 \pi i s_1} \frac{ds_2}{2 \pi i s_2} \frac{(q^{\frac{1}{2}} s_1^{\pm} s_2^{\mp}; q)_{\infty} (q^{\frac{1}{2}} s_2^{\mp}; q)_{\infty}^2}{(q^{\frac{1}{4}} s_1^{\pm} s_2^{\mp}; q)_{\infty}^2}. 
\end{align}
While the boundary condition is quantum mechanically consistent as it is free from the gauge anomaly, 
it is much less clear whether and how such boundary conditions in the ABJM theory 
involving asymmetric couplings of the 2d matter fields to the $U(1)_k\times U(1)_{-k}$ gauge group can be realized in the brane constructions. 

\bibliographystyle{utphys}
\bibliography{ref}

\end{document}